# Diamond phase in space and the possibility of its spectroscopic detection

## A.A. Shiryaev

Frumkin Institute of physical chemistry and electrochemistry RAS,

Leninsky pr. 31 korp. 4., Moscow, 119071, Russia

*The eventual presence of the diamond carbon allotrope in space is discussed in numerous theoretical and experimental studies. The review summarizes the principal mechanisms of nanodiamond formation and experimental results of spectroscopic and structural investigations of nano- and microdiamonds from meteorites. The size dependence of diamond spectroscopic properties is discussed. Infrared spectroscopy allows detection of C-H bonds on surfaces of hot nanodiamond grains. Spectroscopic observation of nitrogen-related point defects in nanodiamonds is very challenging; moreover, such defects have never been observed in nanodiamonds from meteorites. At the same time, photoluminescence and, eventually, absorption of some impurity-related defects, in particular, of the silicon - vacancy (SiV) center, observed in real meteoritic nanodiamonds opens the possibility of diamond detection in astronomical observations.*

## 1. Introduction

The possibility of forming the diamond allotrope of carbon in the interstellar medium was discussed for decades [1-4], but these works have mostly been speculative. A breakthrough occurred in 1987 when nanodiamonds with grain sizes of less than 10 nm were extracted from several types of meteorites in the course of searching for a carrier of isotopically anomalous noble gases [5]. The isotopic composition of trapped noble gases is one of the most remarkable features of these nanodiamonds: it implies that the genesis of at least a fraction of the grains is closely related to explosions of supernova stars of types I [6] or II [5]. Nanodiamond is the first presolar phase available for investigation in a laboratory and is rather abundant [7]. The isotopic composition and structural features of meteoritic nanodiamonds are discussed in numerous reviews, mostly devoted to examining nucleosynthetic processes responsible for creating isotopic anomalies of implanted noble gases [7-12].

Spectroscopic data imply the presence of nanodiamonds in the vicinity of several astrophysical objects, for example, Herbig stars [13]. Such astrophysical studies are often based on a comparison with laboratory studies of synthetic diamonds, which, in turn, differ in many respects from nanodiamonds extracted from meteorites. The diamond phase of carbon in meteorites is represented by grains with widely different sizes ranging from a few nanometers up to hundreds of microns. Mechanisms and conditions of their formation may differ dramatically. In addition to size effects, these differences are reflected in spectroscopic properties. This review discusses the principal properties and spectral peculiarities of real micro- and nanodiamonds from meteorites, the main mechanisms of nanodiamond formation, and their applicability to astrophysical processes. In this review, we will call diamonds with grain sizes smaller than a few ten nanometers nanodiamonds (or grains), and diamonds with sizes from hundreds of nanometers up to microns microdiamonds. The term *macrodiamond* is used to describe still larger crystals, with spectral properties (absorption, luminescence) controlled by lattice defects rather than by surface-related functional groups. In our discussion, the term "size" is applied to the physical dimension of a diamond grain and not of the constituent crystallite.

## 2. Diamonds in meteorites

Diamond samples extracted from meteorites and available for laboratory characterization can be separated into two broad classes with marked differences in size, properties, and, presumably, formation mechanism(s): micro- and nanodiamonds. Emission and absorption bands of intrinsic and impurity-related defects in diamond lattice and of functional groups decorating the surfaces of nanodiamond grains make possible the eventual detection of a diamond form of carbon in space. Peculiarities of the structure and chemical impurities in diamonds from meteorites are discussed in this chapter.

## 2.1. Microdiamonds

Microdiamonds from meteorites share many structural and chemical features with samples produced in a laboratory or on the industrial scale. Although at present there are no reports of astronomical observations of microdiamonds, we discuss some relevant spectroscopic properties of diamond using microdiamond as an illustration.

At the end of the 19th century, diamonds were found in the Novo Urey meteorite, marking the first discovery of extraterrestrial diamond [14]. The respective meteorite class is nowadays termed ureilites, after the impact site. A significant fraction of ureilites contain diamond grains, which may reach several ten microns in size. In most cases, the diamond phase is intimately intergrown with graphitic ($sp^2$) carbon [15]. Despite relatively large dimensions, X-ray diffraction shows that the constituent diamond crystallites (coherently scattering domains) are mostly nanosized, although single crystals up to several microns have recently been observed [16]. The origin of diamond in ureilite remains debatable, but a model of shock transformation of $sp^2$-hybridized carbon in parent bodies of ureilites is the most reasonable. This mechanism is supported by textural graphite±diamond relationships [17-21] and by spectroscopic data on defects and impurities [22]. Carbon forms intermediate between graphite and diamond (diafit) are also found in ureilitic diamonds [23]. Analogous transitional forms were also synthesized at high static pressures (gradia) [24].

Diamonds with a very broad size distribution ranging from nanometers to one micron were found in the enstatite chondrite Abee [25] and in Acfer 182 chondrite [26]. Modeling of their formation is hindered by the isotopic composition of carbon and of impurities (nitrogen and noble gases), which differs markedly from that typical both of ureilitic diamonds and of presolar nanodiamonds. Features of the Abee meteorite matrix are more consistent with a hypothesis of diamond formation during shock transformation of $sp^2$-carbon in the solar nebula [27] than with a model of diamond deposition from the gas phase. The shock mechanism is also supported by the platy morphology of large grains, similar to that encountered in impact diamonds in terrestrial meteorite craters formed by direct shock-induced graphite-diamond conversion [28].

## 2.2. Nanodiamonds

### 2.2.1. Physical properties

The search for the carrier of isotopically anomalous noble gases in meteorites culminated in the discovery of nanodiamonds in chondrites experienced minor degrees of thermal metamorphism (so-called "primitive" chondrites) [5]. It was shown that the (nano)diamond phase (in early papers often called Cd) is the main carrier of noble gases.

According to direct measurements using transmission electron microscopy images, the size of nanodiamond grains falls in the ~1±10 nm range with median at 2.6-2.8 nm [8, 29, 30]. Size distribution of the grains follows a log-normal distribution, consistent with the growth model of their formation [31, 32]. Although the average grain size of nanodiamonds from various meteorites coincides within ~10%, the number of the smallest grains (< 1.5 nm) is inversely proportional to the degree of thermal metamorphism [29]. Analyzing images of the smallest grains is technically difficult, and one might expect an underestimation of their fraction. However, grain mass distribution obtained using MALDI (matrix-assisted laser desorption/ionization) is consistent with electron microscopy results [33, 34]. The differences among meteorites are, presumably, caused by faster destruction (oxidation) of the smaller grains during metamorphism.

Individual nanodiamond grains can be observed directly in the meteorite matrix [35], but searching for them is extremely time-consuming and an overwhelming number of studies are performed using material extracted from meteorites using multi-step treatment with hot acids [36] and/or in a microwave reactor [37]. Direct measurements using electron energy loss (EELS) and X-ray absorption spectroscopy (NEXAFS) showed the presence of carbon with mixed hybridization ($sp^{2\pm x}$) on surfaces of nanodiamond grains, indicating partial graphitization and/or surface reconstruction [38-40]. The grains observed directly in the meteorite matrix are also frequently embedded in graphitic carbon [35, 38, 39]. The origin of this shell remains unclear: it might be genetically linked to nanodiamond formation but may also reflect partial diamond destruction during metamorphism of a meteorite. Macroscopic diamond is resistant to acids; however, partial surface graphitization of the grains and even destruction of the smallest and/or defective grains may occur during thermalmetamorphism of ameteorite and in the course of extraction.

In many papers, the density of meteoritic nanodiamonds is assumed to be ~2.3-2.5 g/cm$^3$, as indicated in pioneering studies [5, 8]. However, such an assumption fully ignores the crucially important point that this value was obtained from the centrifugation behavior of nanodiamond grains covered with nondiamond carbon; actually, this fact was explicitly acknowledged in the initial papers. Consequently, this value is only very indirectly related to nanodiamonds proper; in fact, their density is close to that of macrodiamonds, i.e., 3.51 g/cm$^3$, as follows from the corresponding crystallographic properties.

The most comprehensive investigation of meteoritic nanodiamonds using transmission electron microscopy is paper [29], where the peculiarities of the morphology and internal structure of nanodiamond grains from the Murchison and Allende meteorites are compared to synthetic counterparts. Synthetic samples in that study were represented by nanodiamonds produced by detonation and those grown from the gas phase (CVD - Chemical Vapour Deposition). Comparing the relative abundance of structure features such as various types of twins, dislocations, and stacking faults and related hexagonal diamond polytypes (lonsdaleite and others) led to the important conclusion that nanodiamonds from the studied meteorites are much more similar to grains synthesized from a gas phase than to detonation nanodiamonds. An important feature of the gas-phase nanodiamonds studied in [29] should not be ignored: these samples were obtained in experiments addressing the possibility of homogeneous nucleation of diamond [41]. These experiments are conducted under conditions markedly different from those employed in the synthesis of high quality CVD diamond films (see Section 3.3). Although the studied samples were indeed obtained by deposition from a gas phase, one should not boldly assume similarities in the synthesis conditions between nanodiamonds from meteorites and material obtained in technological processes. Note also that study [29] was done on a bulk sample, which likely represents a mixture of several nanodiamond populations (see Section 2.2.2). Conceivably, systematic investigations of nanodiamond grains with contrasting isotopic compositions may reveal important differences in their formation mechanism.

### 2.2.2. Isotopic composition and chemical impurities

A bulk carbon isotopic composition of meteoritic nanodiamonds is typical of solar system bodies [8, 10-12]. Eventual radial variations in the $^{13}C/^{12}C$ ratio in individual grains might provide information about the grain formation process: for example, enrichment of a grain core in $^{12}C$ and of external layers in $^{13}C$ would imply a contribution of carbon from a He-burning shell of a supernova star [42]. Note, however, that cores of the nanodiamond grains may also be 12C enriched in the case of equilibrium isotopic fractionation [43]. Unfortunately, the small size makes an analysis of the isotopic composition of individual nanodiamonds virtually impossible. Since an average grain contains ~1000 atoms, even random fluctuations of $^{13}C$ atoms (terrestrial abundance of this isotope is ~1.1%) in a specific particle produce significant uncertainties, influencing, for example, the interpretation of atom probe tomography measurements [44], and preclude an unambiguous discussion of eventual isotopic gradients in individual grains.

The nitrogen concentration in nanodiamonds may reach 1±4 at.% [45]; its isotopic composition shows enrichment in $^{14}N$ relative to Earth's atmosphere. At the same time, nitrogen on Jupiter and in the solar wind is, apparently, also isotopically light. Although the nitrogen isotopic composition of different size fractions of nanodiamonds varies, it is still impossible to make sound conclusions about the astrophysical source(s) of this impurity.

Ion implantation of carbon and nitrogen may alter the isotopic composition of pre-existing nanodiamond grains [12]. At the same time, several ten C and/or several N ions should be implanted into a grain to induce significant changes, which, in turn, may lead to amorphization of the grain due to the accumulation of radiation defects. It is thus very difficult to constrain the contribution of the post-growth implantation to the isotopic composition of nanodiamonds.

Mass-spectrometry data provide the main body of information about trace elements in meteoritic nanodiamonds. The extreme sensitivity of the employed methods explains the importance of a detailed consideration of eventual contaminants introduced in the course of sample extraction from a meteorite matrix. Consequently, most attention is directed towards elements with isotopic patterns differing from those observed in the solar system, which allows excluding the contamination issue; other impurities are seldom discussed.

Hydrogen is an important impurity in meteoritic nanodiamonds, with its concentration reaching 20-40 at.% in a bulk sample [46]. It is obvious that the principal part of this impurity is surface bound rather than being incorporated into a crystalline lattice. Surface H-containing groups are discussed in Section 5.1. Hydrogen certainly may enter the lattice in the case of both synthetic [47] and meteoritic nanodiamonds [48], as evidenced by observations of the proton-related H1 defect in their electron paramagnetic resonance (EPR) spectra. For meteoritic nanodiamonds, the distance between neighboring protons of the H1 defects is ~4.6 nm. Recalling that the median grain size is ~2.6-2.8 nm, this value can be interpreted as either the presence of H1 defects in large grains only or the existence of grains devoid of this defect [48].

Hydrogen associated with meteoritic nanodiamonds is enriched with deuterium. Secondary ion mass spectrometry measurements gave values $\delta D_{SMOW} = 180 \pm 11.2$ and $283.7 \pm 51.4$ ‰ [46], which are obviously conservative numbers due to isotopic exchange in the course of sample extraction. The enrichment implies preservation of at least a portion of primordial functional groups. Taking into account the presence of the hydrogen-related H1 defect in meteoritic nanodiamonds, the measured D/H ratio may also include a contribution of hydrogen in the diamond lattice.

As mentioned above, nanodiamonds in meteorites were discovered when hunting for a carrier phase of noble gases. Ion implantation into pre-existing grains is the most probable mechanism of introducing large ions such as noble gases into a diamond lattice [8]. The kinetics of noble gas loss during step heating of nanodiamonds allowed distinguishing several components with contrasting isotopic compositions and release temperatures during pyrolysis [49]. For heavy noble gases (Ar, Kr, Xe), the release pattern is bimodal; for light gases (He, Ne), only one obvious maximum is observed. During pyrolysis, the P3 component is released at low temperatures (between $200 \pm 900$ °C with the maximum at $450 \pm 550$ °C), forming the first peak. The isotopic composition of noble gases of the P3 gases is close, but not identical, to the main reservoir of these elements in the solar system - the Q-component. At higher temperatures (1100±1600 °C with the maximum at 1300 ± 1450 °C), the isotopically anomalous HL component evolves, forming the second peak; at yet higher temperatures, the P6 component with a still poorly determined concentration and composition is observed.

Special attention is attracted by xenon Xe-HL ("heavy and light"), characterized by simultaneous enrichment in the light (the L component) ($^{124, 126}$Xe) isotopes and heavy (the H component) ($^{134, 136}$Xe) isotopes relative to $^{130}$Xe after normalization to the noble gas abundance in the Q-component. The "heavy" component with neutron excess isotopes forms only in the nucleosyntetic r-process; the light isotopes are formed in the p-process. Numerous attempts to physically separate carriers of the L- and H-components failed, implying that they both reside in the same nanodiamond population. Although the composition of Xe-H differs somewhat from predictions of nucleosynthetic models and calls for yet unidentified processes of separation of parent isotopes (see discussion in [11, 12, 44]), the enrichment in r-isotopes is commonly interpreted as a manifestation of a genetic link between nanodiamonds and explosions of type I [6] or II [5] supernovae. With various degrees of confidence, isotopic anomalies are detected both for other noble gases (Kr, Ne, Ar [49]) and for some other elements (Ba and Sr [50], Te and Pd [49], Pt [37, 52]) present in meteoritic nanodiamonds or closely associated with the corresponding fraction of meteoritic material. However, the magnitude of corresponding anomalies is less than for Xe; in addition, a smaller number of isotopes limits the possibility of identifying an astrophysical source(s).

The low temperature of P3 gas release is sometimes explained assuming that this component is surface bound (physisorption or entrapment in the surficial sp$^{2+x}$-hybridized layer) [49, 53]; the high temperature peak (Xe-HL) is assigned to ion-implanted species. However, such an interpretation of the release patterns of Xe and other heavy noble gases is not unique. Ion implantation experiments allowed reproducing themain features of release patterns, such as two peaks for Ar, Kr, Xe and one mode for light gases, even for the implantation of monoenergetic ions [54]. Quantum chemistry modeling allowed proposing an explanation of the observations, showing that the release temperature of implanted ions strongly depends on their atomic configuration in a given grain [55]. Xe ions in the vicinity of cube faces {100} are lost at low temperatures, whereas the ions in the central part of a grain or situated close to octahedral faces

{111} are characterized by two release peaks upon heating. At the same time, the high mobility of implanted helium ions in a diamond lattice leads to a single mode of losses, as is observed experimentally [49].

The lack of isotopically anomalous Xe-HL in the low-temperature release peak implies that nanodiamonds containing this component were annealed prior to entrapment by a meteorite [49]. Since even monoenergetic implantation results in a bimodal evolution pattern, one should consider the contribution of the P3 component also to the high temperature peak [49]. This contribution requires adjustment of the HL-component composition and gives better correspondence to some existing nucleosynthetic models [12].

The discovery of significant deviations (anomalies) of the isotopic composition of oxygen and noble gases in meteorites revealed the presence of a rather broad range of phases of presolar origin (SiC, TiC, graphite, some oxides), genetically linked to novae, AGB stars, and some other astrophysical sources [7]. Identification of presolar grains is based on the detection of isotopic anomalies, and it is possible that the total number of such grains is much higher, but their isotopic composition is close to solar system objects. In some chondrites, the concentration of nanodiamonds may reach 1500 wt.ppm (ppm - parts per million), constituting a significant fraction of the total carbon in meteorites. Thus, ignoring the possibility of "negative selection" of presolar grains (in any case, the magnitude of this effect is impossible to evaluate at the moment), one can conclude that nanodiamonds are the most abundant presolar grains accessible for direct laboratory study. Note that none of the other presolar phases contains HL components of noble gases [7, 12]. Therefore, an apparent contradiction appears: the most abundant (presolar) phase was formed (or genetically linked) during a process (possibly a supernovae explosion), which is not reflected in other materials.

The concentration of noble gases in meteoritic nanodiamonds is relatively high in comparison to most other minerals, but is low in absolute numbers. For helium, the most common element of this group, at most one ion for every 10 grains, is available; there is only one Xe ion for every $\sim 10^6$ grains [47] (more precisely, one should take into account grain size distribution, but too many unknowns preclude this evaluation). Mass-spectrometric studies of C and N isotopic compositions and of nitrogen concentration suggest the existence of several populations of nanodiamonds, differing in both size distribution and the concentration and composition of impurities, which points to different astrophysical sources [12, 56]. The release kinetics of nitrogen and noble gases during stepped pyrolysis and oxidation and analysis of size-separated fractions of nanodiamonds suggest that the grains with Xe-HL are nitrogen-poor in comparison with those containing Xe-P3; in addition, the Xe-P3 grains are larger [12]. Smaller grains are characterized by isotopically heavier carbon. These observations point to different formation conditions (composition of the medium, ion implantation energies) of various populations of nanodiamonds.

Therefore, nanodiamonds extracted from meteorites most likely represent a mixture of several populations, possibly differing in average grain size and astrophysical source. Investigations of one size fraction of nanodiamonds from the Boriskino meteorite reveal the presence of a G-component, related to AGB stars [56]. Whereas for nanodiamonds carrying isotopically anomalous noble gases a genetic link with supernovae explosions is plausible, the origin of nanodiamonds from cometary dust particles [57] and the Abee [25] and Acfer 182 [26] meteorites, and for a fraction, perhaps, considerable, of nanodiamonds from other chondrites, remains debatable, and at present it is unclear whether they are presolar or were formed in the solar system.

### 3. Mechanisms of formation of micro- and nanodiamonds

Figure 1 schematically shows an equilibrium carbon phase diagram and the main regions of diamond synthesis by different methods. Already in 1980s, the possibility of the formation of nanodiamond particles below the graphite-diamond equilibrium line, i.e., in metastable conditions, was shown in thermodynamic calculations and experiments [58-66]. The minimal size and morphology of diamond grains were also investigated in detail [67, 68]. It is now established that nanograins preserve the diamond crystal lattice down to ~1 nm [69]. However, the stability, morphological features, and degree and type of surface reconstruction of grains smaller than 2-3 nm depend not only on size but also on the degree of surface hydrogenation. For small grains (smaller than 2-3 nm), so-called buckydiamond, consisting of a diamond core and a fullerene-like reconstructed shell, is energetically favorable. However, for larger grains, surface reconstruction and (partial) hydrogenation with the eventual formation of $sp^{2+x}$ islands become more probable [68-70]. Such islands (carbon dots) possibly explain some peculiarities of nanodiamond photoluminescence [71]. Depending on the morphology of the nanograin, the fraction of atoms with $sp^{2+x}$ hybridization may reach 25-40 %. Direct measurements supported by quantum chemistry modeling showed the size dependence of grain face curvature: for particles with a diameter of less than 1 nm, the geometrical curvature reaches 1.5 $nm^{-1}$, and the surfaces are no longer flat but are concave [72].

Several of the main methods of formation of diamonds of different sizes are schematically indicated in Fig. 1: (1) HPHT - High Pressures-High Temperatures synthesis under high static PT conditions; (2) synthesis at high dynamic pressures (shock waves); (3) synthesis from the gas phase and/or plasma (CVD/PVD - Chemical/Physical Vapor Deposition); (4) radiation-induced transformation of $sp^2$-carbon into diamond.

### 3.1. Synthesis at high static PT-parameters (HPHT)

Synthesis at high static PT parameters (HPHT) is a well-developed technology, and diamonds with sizes spanning from nanometers up to several grams (more than 100 carats) can be produced. The PT conditions employed cover a very broad field and depend on the growth medium (see Fig. 1); most often, pressures above ~4-5 GPa and temperatures higher than ~900°C are used; a direct graphite-diamond transition requires at least ~12 GPa/ ~1800 °C. Note that in some media nanodiamond nucleation (but not growth!) occurs at considerably lower parameters, starting from 420 °C at 5.5 GPa [70]. In general, at lower thermodynamic parameters, the induction period preceding diamond growth is longer. Natural terrestrial diamonds, with few exceptions, are also formed under high PT conditions, although the growth medium and thermodynamic conditions differ from those in industrial processes [74]. High static PT conditions are sometimes considered in discussions of diamond formation in ureilites [75], but detailed investigations of the structure and defects in such diamonds make this mechanism less plausible; formation in shock waves from collisions of meteorite parent bodies explains most features of ureilitic diamonds [15-22].

### *3.2. Dynamic PT-parameters*

The shock mechanism was also considered in models of nanodiamond formation in interstellar dust. It was suggested that the required *PT* conditions were achieved in collisions of carbonaceous particles [76, 77]. This process may operate in some environments (see below), but it is difficult to explain the formation of a significant number of nanodiamonds due to several reasons. The minimal pressure required for transformation of $sp^2$-hybridised carbon into diamond and the degree of the transformation depend on the structure of the carbonaceous precursor, its density, and the orientation relative to the shock wave direction. The transformation starts at ~18 GPa and is completed at ~33 GPa [78, 79]. Whereas in high quality graphite the transformation is, apparently, martensitic, in disordered carbon, diffusion may play an important role. These factors explain the textural relations of diamond and graphitic carbon on scales down to nanometers and the presence of transitional forms (diafit/gradia) in impact diamonds from meteorite craters and in ureilites [23]. A high abundance of stacking faults, hexagonal diamond polytypes, and dislocations is typical of diamonds formed in dynamic conditions by direct conversion of $sp^2$ carbon and in detonation nanodiamonds (e.g., [80]). At the same time, the concentration of such defects in real nanodiamonds from meteorites is rather small [29]. Furthermore, the size distribution of grains obtained by mechanical destruction usually obeys a power law, whereas, for meteoritic nanodiamonds, a log-normal distribution is observed.

Another important feature of diamond synthesis in dynamic conditions is the necessity of rapid cooling of newly formed grains to prevent graphitization at decreasing pressure. Industrial dynamic synthesis of nano- and microdiamonds involves various means of cooling the explosion chamber, allowing preservation of a significant portion of diamond carbon (e.g., [79, 81]). For example, in detonation synthesis of nanodiamond from a TNT/hexogen mixture at a 40/60 ratio, the following results were obtained: without cooling, the nanodiamond yield is 0.6±3.2 wt.% of the explosive charge, whereas, when a sufficient mass of ice is employed, the yield reaches 10.5 wt.%; the nanodiamond fraction in the explosion product increases from 20-40 to 55-75 wt.% [81]. The density of the charge also strongly influences the nanodiamond yield: excessive porosity leads to the formation of hot spots and resulting graphitization. Radiative losses are the only mechanism of rapid cooling for direct diamond formation in collisions of individual carbonaceous grains in free space. Detailed calculations are needed to evaluate the probability of radiative cooling of diamond nanoparticles from ≥2000 °C (direct graphite-diamond transition) to below 600 °C on timescale of several milliseconds, required for preservation of the diamond phase. Most likely, energetic collisions of carbonaceous grains will predominantly lead to sputtering rather than to the formation of high pressure polymorphs. Nevertheless, nanodiamond synthesis is likely possible in a specific type of collision process. Molecular dynamics modeling shows the possibility of nanodiamond formation in collision of carbon onions with a hard surface [82]. Nanodiamonds form in a very small energy range (~1.5-2 eV on 1 atom, i.e., ~5 km/s); the target should also possess high density. Nanodiamonds may form at higher energies, but excess kinetic energy rapidly destroys them. The efficiency of nanodiamond formation depends, in particular, on details of the atomic structure of carbon layers. These factors limit the applicability of this process.

### 3.3. Deposition from gas phase

Chemical vapour deposition, CVD, was first employed for diamond film growth in 1950-ies [83]. Nowadays, this method is widely used to make nano-, poly- and single crystal diamond films [84-87] and is highly flexible, allowing the use of various molecules as a carbon source and different activation methods (plasma or thermal ionization). Some of employed approaches are actually closer to Physical Vapour Deposition (PVD), i.e. deposition of (diamond) films from ion beams [61, 64] or carbon supersaturated vapour (e.g., [88] and many other).

Gas-phase diamond synthesis is usually performed in a hydrogen or hydrogen-oxygen atmosphere with an admixture of several percent of gaseous hydrocarbons (e.g, $CH_4$, $C_2H_2$) at a heater (or plasma) temperature of 2000-3000 °C and higher; diamond nucleation and growth proceeds on a substrate with temperature of 700-1000 °C [83-87]. Gaseous precursors are activated by electrons (plasma growth) or thermally. Reactions between atomic hydrogen and hydrocarbon molecules lead to the formation of carbon radicals $CH_3$, $CH_2$ and some others; diamond forms upon their deposition on a substrate. Atomic hydrogen or oxygen play a crucial role, destroying long carbon chains in the gas phase, saturating dangling bonds on the diamond surface, and thus preventing surface reconstruction and graphitization; they also selectively remove islands of $sp^2$-hybridised carbon (diamond etch rate is lower).

A comparative study of abundance and type of extended defects (stacking faults, dislocations etc.) described in already mentioned classic paper [29], shows that meteoritic nanodiamonds share many similarities with disperse nanodiamonds obtained by the CVD method. Consequently, gas phase deposition is the most frequently discussed formation mechanism of meteortitic nanodiamonds. Some workers invoke the CVD mechanism to explain the isotopic peculiarities of impurities in ureilitic diamonds [89, 90], but structural and textural properties of these diamonds do not readily fit this scenario.

Unfortunately, in the astrophysical literature, results described in paper [29] are often interpreted too freely without paying attention to important details. Industrial gas phase diamond synthesis is performed in a relatively narrow region of the ternary C-O-H diagram [91, 92]. In many astrophysical papers, it is assumed that diamond formation is possible only within this field, thus strongly constraining the growth models. This opinion is, however, erroneous: the mentioned diagram is valid for the growth of high quality diamond films on a crystalline substrate; at the same time, it is well known that diamond nucleation is possible in a broad range of settings. As an illustration of the difference between growth and nucleation, one may consider deposition of micron-sized diamond grains on sapphire substrate after laser ablation in a pure oxygen atmosphere [93] or deposition of diamond crystals from products of graphite rod evaporation in a hydrogen-free inert gas atmosphere [94]. Therefore, use of the only very limited field of the C-O-H ternary for astrophysical modeling is not reasonable.

Let us also recall that, instead of conventional diamond films, study [29] addressed nano- and microdiamonds obtained in experiments addressing homogeneous diamond nucleation during destruction of chlorine-containing hydrocarbons in a plasma discharge or in hot environments [41]. The conditions of these experiments are very different from those employed in the stationary formation of high quality CVD films. In these studies, hydrocarbons (e.g., acetylene) are irradiated by a $CO_2$ laser or are introduced into a microwave field. In pioneering work, diamond formation was observed only in quench products of laser-induced pyrolysis of gaseous or liquid hydrocarbons, induced by water injection or rapid expansion after a nozzle; in other cases, soot was obtained [95]. Studies of homogeneous diamond nucleation were performed in a broad range of pressures (30-90 mbar for Cl-containing compounds [41], 500-1300 mbar for $C_2H_2$ [96]) at flame temperatures of 500-550 °C and in different atmospheres (nitrogen, argon, $H_2$, $SiH_4$). It was shown that the phase composition, size and purity of diamonds strongly depend of synthesis details, but for every individual experiment, diamond size distribution obeys log-normal distribution. The size of diamond grains varies between a few and several hundred nanometers. Nonequilibrium diamond growth with very high rates (up to $10^6$ μm/s) are important features of these experiments. A high growth rate is, presumably, very important for incorporating high concentration of nitrogen impurity into the lattice as typically observed for meteoritic nanodiamonds (see Section 2.2.2). Remarkably, nanodiamonds were efficiently formed only in experiments with a short residence time of reagents in the flame (2-62.5 ms); at longer exposures (160 ms) diamond does not form [96]. This result is consistent with observations presented in [95]. Graphitisation of just-formed diamond grains due to prolonged exposure to high temperatures is the most likely explanation of this behaviour. An interesting feature of experiments performed in presence of silane is a substantial yield of SiC, likely stemming from temperature rise to at least 700 °C due to high absorption of infrared laser radiation by $SiH_4$ molecules [96]. Recall that SiC grains from meteorites represent of the accessible presolar phase [7].

Taking into account observations described in this section, we conclude that, although deposition from gas phase may be considered the most efficient mechanism of diamond formation in space, conditions required for synthesis of high quality CVD diamonds, such as medium composition and *PT* parameters, should not serve as strict constraints. Rapid drop of temperature of the growth medium is an important factor promoting survival of nanodiamonds.

### *3.4. Diamond synthesis resulting from sp²-C interaction with charged particles and photons*

The possibility of transformation of sp²-C to diamond resulting from irradiation of carbonaceous material by heavy ions or other ionizing particles was discovered during an investigation of uranium-rich shales [97, 98]. Diamond phase formation was confirmed in experiments on heavy ion implantation into graphite [99], irradiation of carbon onions heated to 700-1100 °C with electrons [100] and ions [101]. Internal layers of nanoparticles shrink because of a loss of knocked-out atoms and, in some cases, pressure in nanoparticle core becomes sufficient for stabilization of the diamond phase.

The formation of nanodiamonds as a result of irradiation of carbonaceous matter in meteorites by energetic cosmic rays was suggested in [102]. However, this process fails to explain the observed high concentrations of nanodiamonds in meteorites and isotopic composition of implanted noble gases.

Nanodiamond formation by transforming carbon nanoparticles under irradiation by high-energy photons and ions from a nearby supernova (stochastic heating) was proposed in paper [103]. The synthesis of a small number of nanodiamonds upon ultraviolet (UV) irradiation of compounds mimicking the organic matter of molecular clouds was experimentally realized in work [104]. A mixture of $H_2O$, CO, $NH_3$, and $CH_4$ (4:2:2:1) deposited from a gas phase on an Al plate at 12 K in a vacuum chamber was used as a sample. The integral flux of photons from Ar/Xe/$D_2$ excimer lamps was ~$10^{19}$ ph/cm$^2$. During the irradiation, the substrate was kept at 12 K; all studies of the obtained substances were performed at room temperature. The formation of ~1 nm crystalline grains in an amorphous matrix was shown by transmission electron microscopy. Analysis of interlayer spacing (*d*) revealed graphite (*d* = 0.34 nm) and highly defected diamond with *d* = 0.21 nm (ideally 2.06 nm), termed a "diamond precursor". In addition, crystals with *d* values in the range of 0.21 and 0.34 nm were found and termed diamond-like carbon (DLC). This term is not applicable to results of study [104] since merely by definition DLC – is an *amorphous* matter, eventually with diamond inclusions, and not a crystal with variable interlayer spacing. Possibly, transitional forms such as diafit [23] are present, but require dedicated search.

At the same time, it is known that intense UV irradiation leads to graphitisation of nanodiamonds [105]. Likewise, graphitization and even complete destruction of nanograins during the passage of single ions with sufficient energy was experimentally observed [106] and is caused by overheating of a nanodiamond by the release of energy from the slowing ion [55, 106, 107]. Therefore, the possibility of individual nanodiamond grain formation during irradiation of organic matter is experimentally confirmed. However, the efficiency of this process and, importantly, the survival of formed grains, remain debatable.

## 4. Spectroscopic properties of nanodiamonds

Identification of the diamond form of carbon in astronomical observations becomes possible if phase-specific spectroscopic manifestations of intrinsic and extrinsic defects in a crystal lattice or of some functional surface groups on nanograin surfaces are detected. Infrared (IR) absorption spectroscopy and luminescence are the main methods for investigating point defects in macrodiamonds. Several hundred spectroscopically-active defects are known, most of them involving nitrogen impurities and/or a vacancy in different configurations in the diamond lattice. Other notable impurities, which may be incorporated into the diamond lattice during growth, are boron, hydrogen, and nickel; information about spectral features of a majority of documented defects is available in handbooks [108, 109]. In this review, we discuss only defects (and their spectroscopic manifestations) actually observed in micro- and nanodiamonds from meteorites that could be employed in astronomical studies; other impurities and defects are referred to only briefly.

### 4.1. Infrared spectroscopy

Selection rules forbid one-phonon infrared absorption in an ideal diamond lattice [110]; characteristic lattice bands are observed in the two-phonon region (Fig. 2A). In spectra of bulk nanodiamond materials (e.g., nanocrystalline or impact diamonds) the two-phonon lattice bands are distorted due to changes in some branches of the density of states [47, 111, 112] (Fig. 3). Presumably, these changes are caused by large concentration of extended defects (e.g., stacking faults), but this problem has not studied in sufficient detail. In spectra of disperse and sintered nanodiamonds, the lattice absorption is usually not observed [47].

A comparison with mass-spectrometry shows that, for a macrodiamond, total nitrogen concentration can be evaluated from IR spectra with acceptable precision (e.g., [113]). Nitrogen concentration in natural and synthetic macrodiamonds usually does not exceed 1000 at. ppm [114], sometimes concentrations up to 2000-3000 at. ppm are encountered. Still higher values are almost always associated with nitrogen-containing fluid (gas-liquid) microinclusions [115]. A record concentration of lattice-bound nitrogen is observed in metamorphic microdiamonds from Canada [116], where values up to 9300 ppm (~0.93 at.%) were measured. Although the high concentration of nitrogen impurity in these diamonds is not questioned, considerable discrepancies between IR spectroscopy and mass-spectrometry as well as the small size of the studied crystals [116] complicate a quantitative analysis of lattice nitrogen. According to mass-spectrometry data, nitrogen concentration in meteoritic and synthetic nanodiamonds routinely falls in the 1-4 at.% range [25], i.e. is notably higher than in macrodiamonds. Dedicated studies reveal marked differences between atomic configurations of this impurity in macro- and nanodiamonds. In a macrodiamond, nitrogen mostly substitute carbon in the lattice (see below), in some cases, complexes with vacancies are formed. In a nanodiamond, nitrogen is associated with extended defects such as stacking faults and twin boundaries, and on an atomic scale the nitrogen configuration is dissimilar with that observed in a macrodiamond [40, 117]. Consequently, infrared and luminescence manifestations of nitrogen defects in nano- and macrodiamond differ dramatically.

It is also important to mention that the IR bands of all nitrogen defects in diamond are characterized by a rather complex shape rather than by individual peaks (e.g., [118, 119]; see also Fig. 2b). Therefore, assigning certain maxima in the IR spectra of nanodiamonds to well-known nitrogen defects is, at best, ambiguous. In most cases, absorption bands interpreted as manifestations of nitrogen-related defects in the infrared spectra of meteoritic and of the majority of synthetic disperse nanodiamonds are, in fact, due to surface-bound functional groups.

Based on an extensive dataset for natural and synthetic diamonds, it is currently accepted that, during growth of a macrodiamond, nitrogen is incorporated mainly as a single substitutional atom [113, 119]. This paramagnetic defect gives rise to several bands in one phonon absorption region in infrared spectra and is termed the "C-defect" in notation accepted in IR spectroscopy. In notation employed in the EPR field, this defect is called P1. In synthetic nanodiamonds, a single substitutional nitrogen atom is observed in EPR spectra only and its concentration is low (a few ppm or less) [47, 120].

The unsaturated chemical bond of a nitrogen ion induces significant stress in a diamond lattice [121]; nitrogen diffusion during annealing leads to the formation of a so-called A-defect, consisting of a pair of neighboring substitutional nitrogens. The A-defect is the most common one in natural macrodiamonds. The possibility of incorporating two nitrogen atoms simultaneously (e.g., an $N_2$ molecule) into a growing diamond lattice has been discussed for decades and this may, presumably, indeed happen in some cases ([122] and references therein). However, in the absolute majority of cases, the A-defect forms only during relatively long high temperature annealing, required for overcoming activation barrier (~4.5 eV) of nitrogen diffusion in the diamond lattice.

In some papers, an absorption band with a maximum at 1282 cm$^{-1}$ (7.8 μm) in the IR spectra of synthetic and meteoritic nanodiamonds is interpreted as a manifestation of this defect. Although recent results indicate that A-defects may indeed be present in nanodiamonds, their relative concentration is clearly small and the overwhelming fraction of nitrogen impurity is present in dissimilar forms. Note also that the presence of nitrogen pairs in nanodiamonds is inferred from modeling electron energy loss spectra [123]. However, the calculated spectra of several plausible atomic configurations are comparable, making definite identification difficult.

More complex nitrogen-related complexes - B-defects and so-called platelets - planar exsolutions of interstitial carbon atoms with nitrogen impurities - possess well-defined IR bands with maxima at 1175 (8.51 μm) and 1364-1375 cm$^{-1}$ (7.27-7.33 μm), respectively. Their formation in nanodiamonds appears to be impossible due to yet higher energy barriers (~7±8 eV); moreover, the platelets are several ten nanometers in size and occasionally reach several microns and thus are physically larger than nanodiamond grains.

In an ideal diamond lattice, infrared absorption at the Raman frequency (1332 cm$^{-1}$, 7.507 μm) is symmetry forbidden; however, local distortions due to point defects may relax the constraints [124]. Consequently, many defects induce absorption at this frequency (e.g., A-, B-defects), but only some defects lead to the appearance of a sharp peak. Silicon and transition metals, first of all, Ni and Co, present in some macrodiamonds (e.g., [125]), do not absorb in the IR range and are revealed only indirectly by converting part of the nitrogen to a positively charged state (N$^+$, denoted as X-center). The N$^+$ defect gives rise to a sharp absorption peak at the Raman frequency. Besides the N$^+$ defect, several other centers absorb at the Raman frequency, for example, Y-defects with an as yet uncertain model [126]. This complicates assignment of this absorption line to a particular defect. Despite the rich variety of possible configurations of nitrogen impurities, in nanodiamonds, IR absorption at the Raman frequency is either absent or masked by the contribution of surface functional groups. Paradoxically, Fano resonance leads to the appearance of a "transmission" window near the Raman frequency in spectra of some nanodiamonds (see below).

Boron impurity in a diamond lattice is responsible for several characteristic IR absorption bands [113, 119]; this allows estimating the concentration of uncompensated boron. As in the case of nitrogen, the corresponding bands are not observed in nanodiamonds, although boron doping of the nanograins is certainly possible.

Hydrogen plays an important role amongst lattice impurities in diamond. Besides well-known lines 3107 cm$^{-1}$ and 1405 cm$^{-1}$ (3.218 and 7.117 µm) (Fig. 3, curve 1) plausibly assigned to valence and deformation vibrations in the $N_3HV$ complex [127], numerous IR lines caused by hydrogen-related defects are known in macrodiamonds [128]. In nanodiamonds, IR vibrations of lattice-bound hydrogen are not identified; however, electron paramagnetic resonance spectroscopy reveals defects comprising protons [47, 48]. At the same time, a large number of bands due to H-related surface functional groups are known (e.g., [129]); their composition is determined by thermochemical conditions of extraction and/or treatment and can be altered significantly (Fig. 4) [130]. Vibrations of surface-bound functional groups constitute the most important spectral features of disperse synthetic and meteoritic nanodiamonds [7, 48, 129-138].

Recent studies showed that infrared absorption spectra of nanodiamonds with sizes between 2.6 and 30 nm (possibly also for larger sizes) may contain "transmission windows" in the range of 1100-1460 cm$^{-1}$ (6.85 – 9.1 µm), the highest transmittance is observed at 1328-1332 cm$^{-1}$ (the diamond "Raman" frequency) (Fig. 4, curves 7,8). This spectral feature is due to Fano resonance and points to the existence of a surface conductive shell. However, the exact mechanism of shell formation – surface hydrogenation [132] or the presence of transpolyacetylene fragments and a particular type of surface reconstruction [133] – remains debatable.

### 4.2. Luminescence of nanodiamonds.

The broad featureless photoluminescence band peaked at 5000-6000 Å (Fig. 6, trace 1) is typical of disperse nanodiamonds. Most likely, this band corresponds to intrinsic luminescence of nanosized domains of $sp^2$-hybridised carbon on surfaces of diamond grains; variations ofin the shape, size and structure of these domains are responsible for scatter of spectral properties [71]. The intensity of this band strongly and reversibly changes on heating due to adsorption/desorption of surface-bound water, affecting the probability of radiationless processes [139].

Several hundred luminescing defects in macrodiamonds - both impurity related and intrinsic (vacancies and interstitials)Đare known, and this list is steadily expanding [108, 109, 113, 119]. However, reliable structural models have been developed for a relatively small number of defects; most often, they comprise nitrogen and/or a transition metal (for instance, nickel). Luminescing centers involving elements from group IV of the periodic table are also well known. One of the most studied is the silicon-vacancy (SiV) defect [140]; in recent years, the possibility of incorporating germanium, tin, and lead into a diamond lattice during growth has been demonstrated [141, 142]. Ion implantation permits the incorporation of many chemical elements in diamond; some of them form luminescing defects after annealing radiation defects [143, 144]. In relation to meteoritic nanodiamonds, one may recall that implantation of noble gases may also produce luminescing defects (e.g., [143-146]).

The quantity and quantum yield of luminescing defects markedly drops in nanosized grains. Nitrogen-vacancy (NV) complexes in a neutral or negatively charged state ($NV^0/NV^-$), where V is vacancy) are among the best-studied defects in diamonds of different sizes. Zero-phonon lines of these defects are at 5750 and 6380 Å, respectively [119]. Remarkable optical properties of the NV defect explain the vigorous interest in controlled generation in diamond. It was, however, shown in [147, 148] that the noticeable intensity of NV defect luminescence is observed only for grains larger than ~50 nm; at 4-5 nm - the typical size of detonation nanodiamonds - the probability of their detection becomes negligible. Besides mechanical grinding of macrodiamonds with pre-existing $NV^{-/0}$ defects, irradiation of nanodiamonds is a promising way towards generating these defects in nanograins. Irradiation leads to the formation of vacancies, while diffusion in the course of annealing leads to the formation of complexes with nitrogen. In 4-5-nm nanodiamonds, the $NV^-$ defects appear after high fluence ($n \times 10^{18}$ e$^-$/cm$^2$) irradiation by protons or electrons; however, in order to generate significant concentrations of these centers, high temperature annealing to induce diffusion of vacancies and nitrogen or intense oxidative treatment to remove the distorted surface layer, increasing radiationless losses, is required [149]. The absolute concentration of NV$^-$ defects is rather low: from 1 in ~1000 grains initially up to 1 in ~80-250 grains after irradiation and annealing [150]. The quoted concentrations were measured using electron paramagnetic resonance; the concentrations detected optically are more than an order of magnitude smaller (1 in every ~15,000 grains) [151]. In larger grains obtained by shock compression of nanographite, up to 3 NV$^-$ defects per ~150 nm particle were observed; in finer grains, the concentrations are lower [80]. The spectroscopic properties of nitrogen-vacancy complexes are heavily influenced by luminescence quenching and radiationless processes related to surface functional groups and reconstruction of nanograin surfaces during irradiation.

The size of diamond grains plays is of crucial importance for detecting luminescing defects related to nitrogen or to other impurities. The properties of diamond crystallites several hundred nanometers in size are rather close to macroscopic diamond; for example, stable photoluminescence of SiV and GeV defects in submicron diamonds was reported in [152]. However, in 10 nm and smaller grains only the SiV defect is observed [153-155]; the properties of this center are discussed in some detail in Section 5.2.1. In contrast to nitrogen-vacancy complexes, the SiV center shows photostability irrespective of the grain size [156], although lattice stress and scatter in the defect-surface distance lead to considerable nonhomogeneous broadening of the zero-phonon line [154].

# 5. Spectroscopy of diamonds from meteorites and applications to astrophysical observations

## 5.1. Infrared spectroscopy

As discussed in Section 4.1, absorption by a diamond lattice is not observed in nanodiamonds; absorption by nitrogen defects is also absent. Broad range absorption spectra of nanodiamonds from meteorites are discussed in detail in references [8, 134-138]. As expected, the absolute majority of spectral features in the IR region are caused by various functional groups on the surfaces of nanodiamond grains. The bands can be observed both in absorption spectra and in the emission of heated grains. Unfortunately, in many papers discussing infrared spectra of meteoritic nanodiamonds, some of the observed bands are assigned erroneously. Primarily, the main faults are related to the interpretation of absorption bands in the range of 1000-1500 cm$^{-1}$ (6.6-10 µm) as manifestations of nitrogen-related defects. The incorrectness of this assignment is discussed in Section 4.1 and is related to the absence of nitrogen-related bands in nanodiamond spectra.

Extracting nanodiamonds from a meteorite involves dissolution of the matrix (mainly silicates and oxides) in acids and inevitably leads to alteration of the physical-chemical state of the grain surface and speciation of functional groups. Enrichment of nanodiamonds with deuterium (Section 2.2.2) indicates that a fraction of the primordial hydrogen survives the extraction. However, since a specific protocol of the thermochemical treatment is reflected in the infrared spectra of nanodiamonds [134, 136], the composition of surface functional groups on pristine nanodiamonds and the applicability of their spectroscopic detection for astrophysical purposes remain uncertain and require additional studies.

Spectroscopic investigation of macrodiamonds without proper surface cleaning from grease (literally, fingerprints), frequently reveal vibrations of sp$^3$ C-H bonds in methyl groups in the range 3.4-3.5 µm (Fig. 3, trace 2); similar bands are common for CVD diamond films. For macrodiamonds, these lines are almost always due to surface contamination and are of little interest; in the CVD case they might be inherited from growth processes. Studies of synthetic nanodiamonds showed a size dependence of intensities and fine structure of the C-H bands: whereas, for grains larger than 50 nm, a doublet of sharp lines at 3.43 (~2915 cm$^{-1}$) and 3.53 µm (~2832 cm$^{-1}$) with a complex envelope is observed, for smaller grains the fine structure becomes less pronounced, the main maxima at 3.43 and 3.53 µm broaden and the gap between them becomes shallower [138]. These absorption bands correspond to vibrations of sp$^3$ C-H- groups on {100}, {110} and {111} faces of diamond crystals [157, 158]. The size dependence of the spectral bands is caused by the decreasing area of the flat faces at diminishing grain size. The C-H bands at 3.43 and 3.53 µm are also observed for nanodiamonds from meteorites [138]. Importantly, nanodiamonds studied in [138] were heated in a vacuum and treated in hydrogen plasma after the extraction from a meteorite. Thus, obtained results are of interest for physical chemistry of very small diamond grains (synthetic nanodiamonds are mostly larger), but, strictly speaking, are not very useful for investigating primordial surface chemistry of meteoritic nanodiamonds.

From the astrophysics point of view, the interest to the C-H vibrations at 3.43 and 3.53 μm is due to their detection in spectra of protoplanetary disks of Herbig stars Ae/Be HD 97048, Elias 1 and post-AGB star HR 4049 [13, 159]. The mentioned features, in particular, the 3.5 μm one, were observed in spectra of these stars a long time ago [160-163]. Based on laboratory studies, it was suggested that emission lines in vicinity of HD 97048 are due to formaldehyde complexes [164], however, later, a connection of this band with grains heated to >1000 K was found [165]. In a groundbreaking paper [13] the close correspondence of spectral features of the Elias 1 and HD 97048 with spectra of hydrogenated diamonds was shown. Minor spectral shift can be explained by the relatively high temperature of carrier grains. Analysis of spectra of 82 Herbig stars showed the presence of the 3.53 μm feature in three objects only (4% of the set), and the 3.43 μm line was observed in two cases only (i.e. ~2%) [166]. No obvious correlation between age and/or other spectral properties of a central star and probability of nanodiamond lines detection was found. An exception is the weak correlation between integral luminosity of the central star and nanodiamond lines, which can be due to stronger heating of the emitting grains.

Spatially-resolved observations of objects with "nanodiamond" lines showed that these features are confined to internal parts of dust disks (distance from central star: less than 15 a.u.), whereas polyaromatic compounds and silicate grains emit in outer parts [159, 167-169]. Such a spatial distribution is not fully understood. It might be explained by the formation of nanodiamonds when polyaromatics and silicates are already exhausted [159], or in the framework of diamond formation by irradiation of carbon onions (see Section 3.4) from X-ray flares of the central star, assuming that in the outer disk part the transformation process is not yet completed and emission of nanodiamond cores is masked by a graphitic shell [169]. The absence of the nanodiamond signal in the peripheral parts may also be explained by recalling that the mentioned emission bands possess significant intensity only in the case of relatively hot (800±1000 K) nanodiamonds. Therefore, nanodiamonds could also be present in the outer disk parts, but their temperature is too low to allow observations in the IR range. Another plausible scenario, which does not contradict the previous one, suggests that the size distribution of nanodiamonds in the disk changes in the radial direction and, consequently, the efficiency of their heating by the stellar radiation changes. Noticeable differences in photoluminescence spectra of nanodiamonds extracted from meteorites of various petrological classes [170] could stem from the existence of such radial heterogeneity in the protoplanetary disk of the early solar system.

Models constructed from calculations of radiation losses suggest that the size of nanodiamond grains responsible for the IR emission in vicinities of HD 97048 and Elias 1 falls in 1-10 nm range [159]. Despite the generally meticulous approach used in that paper, some of the employed parameters were selected erroneously; for example, following Ref. [5], the diamond density was taken as 2.3 $g/cm^3$; the incorrectness of this value was shown above. However, correcting the density would make a rather minor contribution to the calculated size distribution. At the same time, spectral peculiarities of the C-H vibrations in these astrophysical objects point to grain size of at least 25 nm. Ignoring eventual mistakes in modeling of the radiation losses, the discrepancy between the calculated and inferred sizes of nanodiamonds may be explained by assuming incomplete hydrogenation of the grains. Interestingly, quantum chemistry modeling points to instability of nanodiamond particles with bare surfaces and suggests that they delaminate to onion-like shells. However, (partial) surface hydrogenation stabilizes the diamond structure [67, 70].

### 5.2 Luminescence

In line with investigation of IR absorption and emission spectra, luminescence spectroscopy represents another promising approach to the observation of nanodiamonds. Moreover, to date, luminescing point defects have been the only actually recorded spectroscopic manifestation of meteoritic nanodiamonds themselves and not of their surface states [40, 154, 170].

Luminescence of diamonds from ureilites is similar to that of macrodiamonds and is dominated by NV and H3 complexes (pair of substitutional nitrogen atoms separated by a vacancy) [22, 171] (Fig. 5). Nitrogen in ureilitic diamonds is inherited from a carbonaceous precursor. The luminescing defects presumably form during annealing of the enveloping meteorite, for example, after strong shocks, evidenced by numerous textural and structural features. The formation of these defects directly during the impact is also possible and is assisted by enhanced diffusion of nitrogen and vacancies during shock-induced temperature excursions [172].

Spectra of the induced thermoluminescence of diamonds from several meteorites are reported in Refs. [173, 174]. For all nanodiamonds, a gradual increase of thermoluminescence intensity with temperature is observed; a weak peak with a sample-dependent maximum position between 280-330 °C is present. At the same time, spectra of diamonds from Novo Urey ureilite and Abee chondrite show pronounced peaks at ~290 and ~130/320 °C, respectively. The gradual intensity rise with temperature may be related to surface $sp^2$-C and the peaks in all samples correspond to a common class of defects. The glow curve of the last two samples are similar to those reported for irradiated CVD diamond films [175, 176], but available data do not allow an unambiguous interpretation of the experimental results relating the glow peaks to known defects.

Photoluminescence spectra of meteoritic and disperse synthetic nanodiamonds are generally similar and usually consist of a broad band with maximum at 5500-6500 Å (Fig. 6). As discussed in Section 4.2, this band may be related to islands of carbon with $sp^{2+x}$ hybridisation [71]. Spectral envelope often show some features and inflections, but their assignment is very ambiguous. Importantly, a large breadth of the band is preserved upon cooling of the sample (nonhomogeneous broadening). The general shape of the typical nanodiamond spectrum is similar to the emission of G9-K0 stars and thus cannot be considered a diagnostic feature. At the same time, luminescence of point defects can open a promising approach to the detection of nanodiamonds in astrophysical studies. However, a direct correspondence between luminescent lines known for defects in macrodiamonds and spectra of nanoparticles is most often erroneous due to both drastically different atomic configurations of defects in diamonds of various sizes and large surface contribution to nonradiative processes. For example, in Refs [2, 171], a large number of sharp lines in spectra of Diffuse Interstellar Bands are interpreted as manifestations of nitrogen-related defects in diamond, excited by photons and electrons from the stellar wind. However, all considered defects form in diamond macrocrystals only after prolonged high temperature annealing at high stabilizing pressures.

### 5.2.1 Luminescence of point defects

In papers [178, 179] spectra of Extended Red Emission (ERE) of several galaxies were assigned to luminescence of nitrogen-vacancy (NV) defects in ~100 nm diamond grains. Although this assignment might seem reasonable, several important considerations makes it rather implausible. First, as discussed in detail in Section 4.2, the mere observation of the NV defect in nanodiamonds is extremely difficult. In experimentally studied nanodiamonds from meteorites, these defects are not observed at all, probably, due to the small grain size, responsible for the decreased probability of NV center formation and its vanishing intensity [147, 148]. Second, whereas experimental luminescence spectroscopy usually employs a narrow bandwidth laser line, the excitation spectrum of a star is closer to a continuum. In this case different charge states of the $NV^{0/-}$ defect would be observed and a resulting spectrum will be significantly broader than ERE. Note, that the spectral envelope of nanodiamonds studied in Ref. [178] changes markedly with excitation and the correspondence with spectra of particular astrophysical objects is achieved only with narrow bandwidth excitation. Finally, a comparison of maximum position and half width of ERE for very different objects, such as various nebulae, HII regions and galactic halo [180], with data for real meteoritic nanodiamonds shows obvious dissimilarity: the nanodiamonds' emission band is 5-10 times broader than that of ERE with a similar maximum position.

At present, the only luminescing defect seen in real nanodiamonds from meteorites is the negatively charged defect silicon vacancy (SiVÿ) [40, 154, 170]. This defect is observable not only in emission but also in absorption as a band peaking at 7370 Å [140]. In spectra of this defect in meteoritic nanodiamonds, a phonon replica at ~7510 Å (Fig. 6) and, possibly, a weak feature at ~7200 Å are also detected [40]. At room temperature, the half-width for a bulk sample (averaged for many grains) is ~120 Å, and for individual grains, ~40 Å. The band maximum is shifted ~18 Å (0.004±0.007 eV) towards higher energies relative to the SiV center in perfect diamond film [154]. The shift and sample-dependent scatter in the peak position are likely associated with significant strain in individual nanograins and variations in distances between the defect and the surface. In this connection, results of modeling of the influence of nanodiamond grain morphology and surface reconstruction on elastic properties are of interest [181]. The observed peculiarities of luminescence of individual grains are possibly related to their morphology.

Investigation of samples from several meteorites showed that the SiV defects are encountered mostly, if not exclusively, in the smallest nanodiamonds with grain sizes of less than 2 nm, comprising fewer than 600 atoms [154, 170]. Clearly, we discuss here silicon impurity in diamond lattices and not micro- and nanosized SiC grains present in meteorites and often observed in insufficiently purified nanodiamond material [7]. The mechanism of incorporation of silicon ions into meteoritic nanodiamonds remains unclear, since this impurity easily enters growing diamonds and can also be driven in by ion implantation.

The strong temperature dependence of the luminescence intensity of this defect constrains the temperature of emitting grains to less than ~450 K. The SiV luminescence is characterized by a relatively high quantum yield, a narrow width of the zero-phonon line, and a rather flat excitation curve in a broad spectral range [182]. Excitation with UV photons with energies close to the band gap width is very efficient. These features make SiV luminescence more promising in the search for nanodiamonds than that of nitrogen-vacancy complexes. Note that irradiation of diamond with electrons or X-rays may convert some SiV$^-$ defects to the neutral charge state (SiV$^0$), giving rise to luminescing center 9464 Å [183].

One remarkable feature of this impurity is the presence of up to 4 silicon atoms per nanodiamond grain [154], which makes it the third most abundant impurity after nitrogen and hydrogen. The isotopic effect may serve as yet another explanation for the shift of the luminescing band from that in macrodiamond. Unfortunately, our own attempts to directly measure silicon isotopic composition in meteoritic nanodiamonds using accelerator mass spectrometry failed due to sample contamination during extraction and difficulties in finding a matrix with a silicon concentration below the detection limit of the method.

Significant interest in understanding the behavior of noble gases in meteoritic nanodiamonds and a very high probability of an implantation mechanism of incorporation suggest that the detection of corresponding luminescing centers be attempted. Such centers are known for helium and neon (He, Ne) [143-146, 184], but up to now they have not been observed in meteoritic samples. In most cases, annealing is required to convert a portion of implanted ions to a spectroscopically active configuration. It is important to note that the process of heavy ion implantation itself may lead to significant heating of a nanodiamond grain [55, 106], but this temperature rise is, presumably, insufficient to form stable luminescing centers. This hypothesis is supported by experiments on the implantation of Kr and Xe ions in nanodiamonds: although incorporation of these ions into a diamond lattice is confirmed by X-ray absorption spectroscopy, no luminescence of the corresponding centers were found [185]. This fact may, in fact, be explained by the complex dependence of the stability of configurations of implanted ions on the distance to grain faces with different crystallographic indices.

## 6. Possible astrophysical sources of nanodiamonds

Although meteoritic nanodiamonds have been studied for several decades, their astrophysical source(s) and formation mechanism are still not known unambiguously. The main body of data is available on the isotopic properties of carbon and of impurities in the nanograins. There is little doubt that nanodiamonds extracted from meteorites represent a mixture of several populations, likely genetically linked to diverse astrophysical sources. Formation mechanisms of nanodiamonds of these populations may also differ.

As discussed in Section 3, the diamond phase of carbon can be formed in very different processes. A "bottleneck" in discussions of the genesis of meteoritic nanodiamonds is the lack of detailed and representative studies of their structure, with the notable exception of [29], where deposition from the gas phase was proposed. In particular, the extensive set of experimental data obtained after publication of this seminal paper suggests that nanodiamonds of differing geneses may vary in size. The smallest grains may structurally differ from the larger ones, implying different formation mechanisms. However, these proposals require experimental verification, which is not yet available.

The metastability of diamond exceeding several nanometers in size implies the operation of non-equilibrium processes during gas-phase deposition. The necessity of cooling just formed grains to temperatures less than ~300-400 °C on a time scale up to several tens milliseconds to prevent graphitization is an extremely important requirement. Seed particles were not observed in studies using transmission electron microscopy, implying homogeneous nucleation. In all studies of homogeneous nanodiamond nucleation, very high growth rates unusual for conventional CVD synthesis were reported. In turn, this points to fairly high concentration of carbonaceous precursors. Modeling of dust formation in the vicinity of supernovae and carbon stars shows that required densities can be achieved only in dust-gas clumps [186, 187]; shock waves may additionally trigger dust formation.

Impurities also point to possible links between diamond formation and dust-gas clumps. It is known that the efficiency of nitrogen incorporation into diamond during laboratory gas phase synthesis is low due to the small probability of destroying molecular nitrogen due to the insufficient energetics of the employed plasma. Introducing compounds with weakly bound nitrogen, such as $NH_3$, increases the concentration of this impurity in the film; similar processes operate during growth at high static parameters (see discussion in [40]). High growth rates are, presumably, necessary for the incorporation of high nitrogen concentration in a nanodiamond lattice. Therefore, if the hypothesis of diamond formation in a dust-gas clump is correct, this implies the necessity of the presence of compounds with weakly bound or atomic nitrogen, in addition to carbonaceous precursors [40]. The emergence of nitrogen available for incorporation into a diamond may also result from the passage of a shock wave.

Diamond formation by irradiation of onion-like carbons does not require high quench rates but is feasible only in an intense stream of charged particles: diamond forms in onions at doses of ~600 displacements per atom [101]. The energy of the particles should exceed the displacement energy of target atoms, i.e., 20-25 eV for graphite. Although diamond is resistant to radiation, the accumulation of defects results in structure distortions and eventually to amorphization. In Ref. [169], the formation of nanodiamonds in the vicinity of the star Elias 1 was linked to powerful X-ray flares. The random character of irradiation of nebular material by the flares may, presumably, enable the formation and preservation of nanodiamonds. However, this mechanism cannot be responsible for the formation of nanodiamonds with a high concentration of nitrogen impurity.

According to isotopic data, several possible sources of nanodiamonds could be proposed: supernovae stars and interstellar matter processed by explosion shock waves, carbon stars, or the solar system.

### 6.1. Supernovae

The contribution of supernovae stems from isotopically anomalous noble gases, Te, and some other elements. After the discovery of nanodiamonds in meteorites, precisely supernovae were considered as the main source, and the isotopic composition of carbon and nitrogen were also explained by models of turbulent mixing of supernovae II shells [42]. Subsequent studies showed that the population of nanodiamonds with isotopically anomalous xenon is depleted in nitrogen [12], and the applicability of early models is questioned. Existing models of nucleosynthesis in supernovae do not explain all observed peculiarities of noble gas abundances. The best fit is obtained in a model assuming physical separation of stable isotopes of Xe from radioactive, but not yet decayed, parent isotopes of I and Te on a timescale of approximately 2 hours after the explosion [188]. Taking into account that the bimodality of the release pattern of heavy noble gases is explained by peculiarities of the implantation process [54, 55], the isotopic composition of noble gases, especially the heavy component Xe-H and isotopic ratio of Kr and Te, may be described in the framework of the fast separation model [12]. The light component may also be explicated within this model, but it requires serious constraints on the pathways of formation of certain isotopes. Therefore, the fast separation model explains many features of the isotopic composition of nanodiamonds, but a mechanism of this separation has not been proposed. Kilonova events were also proposed as possible sources of nanodiamonds [44], but existing nucleosynthetic models of such events lack the precision required for a comparison with experimental data.

In any case, it is unclear why isotopically anomalous noble gases are not encountered in other phases, such as SiC and graphite, for which genetic links to supernovae are plausible. One of the considered models suggests spatial segregation of regions where the phases are formed: graphite and SiC originate from carbon-rich expanding remnants of the He shell, whereas nanodiamonds form in the hydrogen shell [44]. However, the mechanism of Xe-HL transport from the deep stellar interior directly to the H shell, bypassing the He layer, is unspecified.

The formation of nanodiamonds in the vicinity of a chemically peculiar star seems to be a worthwhile hypothesis addressing an explanation of isotopically anomalous xenon. Very high Xe concentrations were measured in many HgMn and B-type stars (e.g., [189, 190]). At present, there are no hints of isotopic stratification of this element, possibly due to the small isotopic shift of Xe II spectroscopic lines [191]. If, however, the stratification exists, for example, for other charge states, it might explain isotopically anomalous xenon released from meteoritic nanodiamonds.

Most likely, the population of nanodiamonds related to supernovae explosions is relatively small and other astrophysical sources play a more important role. The absence of Xe-HL in the low temperature gas release peak implies that the supernova-linked population of nanodiamonds was heated to at least 600 °C. Apparently, these grains are, on average, smaller than the median size of all nanodiamonds and, as noted earlier, are depleted in nitrogen. Interestingly, the luminescing SiV defect is associated with small grains. Unfortunately, at present, it is unclear whether the correlation between Xe-HL and SiV is a simple coincidence or is genetically meaningful.

### 6.2. Carbon stars

Spectroscopic detection of nanodiamonds in surroundings of carbon star HR 4049 [159] is consistent with presence of Xe $s$-isotopes, neon, and heavy carbon in some nanodiamond fractions [56]. Directly after formation these grains will lack Xe-HL. Implantation of gases is possible either if the carbon star forms a binary and the second component explodes as type I supernova, or in interstellar medium [6]. Nucleation of nanodiamonds around binary Wolf-Rayet stars characterized by periodic dust formation [192] is a variant of such scenario. For implantation into a nanodiamond grain relative velocity of Xe ions does not exceed 40 km/s. These conditions could be met in collisions of a shock wave from exploded star with dust clouds expelled at earlier stages of evolution; moreover, this mechanism explains well the necessity of rapid formation of nanodiamonds with high nitrogen content.

### 6.3. Solar system

The formation of some nanodiamonds directly in the solar system and their mixing with presolar grains is consistent with bulk carbon and nitrogen isotopic composition. The P3 component of noble gases differs marginally from those typical for solar system objects.

Possibly, the implantation of the P3 gases predated full homogenization of protosolar nebula or occurred at a pre-main sequence stage of the Sun's evolution. The detection of nanodiamonds in surroundings of the Herbig stars HD97048 and Elias [13, 159] fits this scenario.

## 7. Conclusions

Infrared and luminescence spectroscopies are the most promising approaches to detecting diamond carbon in space. Ultraviolet spectra do not seem suitable due to the large width of spectral features and ambiguity in assignment.

At present, known infrared signatures of nanodiamonds are related to well-defined configurations of hydrogen-containing functional groups on the surfaces of relatively large grains with sizes of at least 25-100 nm. Currently, only thermal emission of nanodiamond grains heated to >800-1000 K is being detected, but cold nanograins may, in principle, be observed in absorption spectra. However, for typical meteoritic nanodiamonds available for characterization in the laboratory, imperfections in the crystalline lattice and distortion of shapes of the smallest grains lead to broadening of surface C-H vibrations and smearing of the features. Subsequently, infrared spectroscopy makes possible reliable detection of heated, relatively large diamond grains, and is less suitable for smaller ones. Assignment of absorptions in the 8-20 mm region to nitrogen defects in diamond lattice is erroneous.

Luminescence spectroscopy is an independent method of detection of nanodiamonds with temperatures less than ~450 K. At present, the only luminescing defect in meteoritic nanodiamonds is the silicon-vacancy complex (SiV), manifested as a line peaking at 7370 Å (Fig. 6) with a broad excitation spectrum. Irradiation may change the charge state of the defect and lead to the appearance of a luminescence peak at 9460 Å. Both defects are also observable in absorption. The SiV defect luminescence appears to be promising for the detection of nanodiamonds in astronomical observations due to several reasons: (1) this defect is definitely present in real nanodiamonds from meteorites; (2) the zero-phonon line is narrow even in nanoparticles; (3) the luminescence intensity is only weakly dependent on excitation energy and possesses a high quantum yield.

The search for nanodiamonds aimed at detecting luminescence of nitrogen-vacancy complexes ($NV^{0/-}$) is unlikely to be successful due to important complications. The main difficulties are the strong dependence of spectroscopic properties of these centers on grain size and excitation conditions, its problematic observation in grains of less than 25-50 nm, and, importantly, its absence in studied meteoritic nanodiamonds. Moreover, the emission band of these defects (5700-6400 Å) is very broad and overlaps with luminescence from many other substances.

Detecting nanodiamonds using luminescence lines of XeV (7930-7940 and 8110-8130 Å [184]) is hindered by very low concentrations of implanted Xe ions. The observation of implanted helium ions (5225, 5365, 5605 Å) and/or neon (7160 Å) [145, 146] may be more realistic. Information about optically active manifestations of other elements found in nanodiamonds using mass-spectrometry and spectroscopy (H, Ar, Kr, Pd, Sr, Te, Ba) is still unavailable.

**Acknowledgements**

This work was supported by Ministry of Science and higher education of Russian Federation (№ 122011300052-1). I am grateful to U. Ott and A.B. Verkhovsky for sharing some experimental data prior to publication.

**References**

1. Saslaw W, Gaustad J. *Nature* **221** 160 (1969)

2. Duley W W *Astrophys Space Sci.* **150** 387 (1988)

3. Allamandola L J et al. *Astrophys. J.* **399** 134 (1992)

4. Allamandola L J et al. *Science* **260** 64 (1993)

5. Lewis R S et al. *Nature* **326** 160 (1987)

6. Jorgensen U G *Nature* **332** 702 (1988)

7. Anders E, Zinner E *Meteoritics* **28** 490 (1993)

8. Lewis R S, Anders E, Draine B T *Nature* **339** 117 (1989)

9. Daulton T L in *Ultrananocrystalline diamond* (Eds Shenderova O, Gruen D M) (Norwich: William-Andrew, 2006) p. 23

10. Ott U *Space Sci. Rev.* **130** 87 (2007)

11. Ott U *Chem Erde-Geochem.* **74** 519 (2014)

12. Ott U, Verchovsky A B, Daulton T L in *Presolar Grains in Extraterrestrial Materials: Probing Stars with Stardust* (Ed. Amari S) (Elsevier, 2024)

13. Guillois O, Ledoux G, Reynaud C *Astrophys. J.* **521** L133 (1999)

14. Erofeev M V, Lachinov P A *Zap. Imper. Miner. Ob-va, SPb*, **24(2)** 263 (1888); Jerofejeff M, Latschinoff P *Verh. Russ. Kais. Min. Gesellsch. St. Petersburg* **24**(2) 263 (1888)

15. Vdovykin G P, *Space Sci. Rev.* **10** 483 (1970)

16. Nestola F et al., *PNAS* **117(41)** 25310 (2020)

17. Valter A A et al,. *Geokhimiya* **33**(10), 1027 (2003); Valter A A et al,. *Geochem. Int.* **41**(10) 939-946 (2003)

18. Lipschutz M E *Science* **143** 1431 (1964)

19. Nakamuta Y, Toh S *Amer. Miner.* **98** 574 (2013)

20. Garvie L A G, Németh P, Buseck P R *Amer. Miner.* **99** 531 (2014)

21. Nakamuta Y, Kitajima F, Shimada K *J. Miner. Petrol. Sci.* **111** 252 (2016)

22. Lorenz C A et al. *Meteorit. Planet. Sci.* **54**(6) 1197 (2019)

23. Nemeth P et al. *PNAS* **119(30)** e2203672119 (2022)

24. Luo K et al. *Nature* **607** 486 (2022)

25. Russell S S et al. *Science* **256**(5054) 206 (1992)

26. Grady M M et al. *Earth Planet. Sci. Lett.* **136** 677 (1995)

27. Rubin A E, Scott E R D *Geochim. Cosmochim. Acta* **61**(2) 425 (1997)

28. Masaitis V L, Mashchak M S, Raikhlin A I, Selivanovskaya T V, Shafranovskii G I *Almazonosnie impaktity Popigaiskoi astroblemy* (*Diamondiferous impactites of Popigai astrobleme*) (SPb, VSEGEI) 1998; *Popigai Impact Structure and its Diamond-Bearing Rocks* (Ed. V L Masaitis) (Springer Cham, 2019)

29. Daulton T L et al. *Geochim. Cosmochim. Acta* **60** 4853 (1996)

30. Fraundorf P et al. *Ultramicroscopy* **27** 401 (1989)

31. Eberl D D, Drits V A, Srodon J *Am. J. Sci.*, **298(6)**, 499 (1998)

32. Eberl D D *Amer. Miner.*;**109(1)**, 2 (2024)

33. Lyon I C *Meteorit. Planet. Sci.* **40**(7) 981 (2005)

34. Maul J et al. *Phys. Rev. B* **72** 245401 (2005)

35. Banhart F. et al., *Meteorit. Planet. Sci.* **33 Suppl.**, A12 (1998)

36. Tang M et al. *Geochim. Cosmochim. Acta* **52** 1221 (1988)

37. Merchel S et al. *Geochim. Cosmochim. Acta* **67**(24) 4949 (2003)

38. Garvie L A J, Buseck P R *Meteorit. Planet. Sci.* **41**(4) 633 (2006)

39. Garvie L A J *Meteorit. Planet. Sci.* **41**(5) 667 (2006)

40. Shiryaev A A et al. *Geochim. Cosmochim. Acta* **75** 3155 (2011)

41. Frenklach M et al. *Appl. Phys. Lett.* **59** 546 (1991)

42. Clayton D D et al. *Astrophys. J* **447** 894 (1995)

43. Shiryaev A A et al. *Phys. Chem. Chem. Phys.* **22** 13261 (2020)

44. Lewis J B et al. *Meteorit. Planet. Sci.* **55** 1382 (2020)

45. Russel S S, Arden J W, Pillinger C T *Meteorit. Planet. Sci.* **31** 343 (1996)

46. Virag A et al. 1158 *Lunar Planet. Sci. Conf.* XX (1989)

47. Shiryaev A A et al. *J Phys. Cond. Mat.* **18** L493 (2006)

48. Braatz A et al. *Meteorit. Planet. Sci.* **35** 75 (2000)

49. Huss G R, Lewis R S *Meteoritics* **29** 791 (1994)

50. Lewis R S, Huss G R, Lugmair G. *Lunar Planet. Sci. Conf.* XXII, 807 (1991)

51. Maas R. et al. *Meteorit. Planet. Sci.* **36**, 849 (2001)

52. Wallner A. et al. *Nucl. Instrum. Meth. Phys. Res. B* **294**, 496 (2013).

53. Fisenko A V, Semjonova L F *Geokhimiya* **48**(12), 1257 (2010); Fisenko, A.V., Semjonova, L.F. *Geochem. Int.* **48(12)**, 1177–1184 (2010).

54. Koscheev A P, et al. *Nature* **412**, 615-617 (2001)

55. Aghajamali A, Shiryaev A A, Marks N A *Astrophys. J* **916** 85 (2021)

56. Verchovsky A B et al. *Astrophys. J* **651** 481 (2006)

57. Dai Z R et al *Nature* **418** 157 (2002)

58. Fedoseev D V et al., *Pis'ma v ZhETF* **32**(1) 7 (1980); Fedoseev D V et al., *JETP Lett.* **32(1)**, 5 (1980).

59. Fedoseev D V, Varnin V P, Deryagin B V, *Uspekhi khimii* **53**(5) 753 (1981); Fedoseev D V, Varnin V P, Deryagin B V *Russ. Chem. Rev.*, **53**(5), 435 (1984)

60. Fedoseev D V *Carbon* **21**(3) 237 (1983)

61. Chaikovskii E F, Puzikov V M, Semenov A V, *Kristallografia* **26**(1) 219 (1981); Chaikovskii E F, Puzikov V M, Semenov A V, *Soviet Phys. Crystallogr.* **26**, 122 (1981)

62. Chaikovskii E F, Rozenebrg G Kh, *Doklady AN SSSR* **279** 1372 (1984); Chaikovskii E' F, Rozenberg G Kh *Sov. Phys. Dokl.* **29** 1043 (1984)

63. Chaikovskii E F et al. *Doklady AN Ukr.SSR*, A, №11 53 (1985)

64. Chaikovskii E F et al. *Surface. Fizika, Khimiya, mekhanika,* **9** 98 (1985)

65. Gubin S A et al. *Khimicheskaya fizika* **9**(3) 401 (1990); Gubin S A et al. *Sov. J. Chem Phys.* **8(3)** 662 (1991)

66. Gutzow I et al. in *Nucleation Theory and Applications* (Ed J W P Schmelzer) (Wiley-VCH Verlag GmbH & Co, KGaA) 2005

67. Raty J-Y, Galli G, *Nat. Mater.* **2** 792 (2003)

68. Barnard A S, Russo S P, Snook I K, *J.Chem.Phys.* **118(11)** 5094 (2003)

69. Stehlik S et al., *Sci.Reps.* **6** 38419 (2016)

70. Ekimov E A et al., *Diam.Relat.Mater.* **136** 109907 (2023)

71. Shenderova et al., *Part. Part. Syst. Charact.* **31** 580 (2014)

72. Chang S L Y et al., *Nanoscale Horiz.* **3** 213 (2018)

73. Bundy F P et al., *Carbon* **34(2)** 141 (1994)

74. Luth R W, Palyanov Y N, Bureau H, *Rev. Miner. Geochem.,* **88** 755 (2022)

75. Urey H C, Mele A, Mayeda T *Geochim. Cosmochim. Acta* **15** 1 (1957)

76. Tielens A et al. *Astrophys. J* **319** L109 (1987)

77. Blake D F et al. *Nature* **332** 611 (1988)

78. Pyaternev S V, Pershin S V, Dremin A N *Fiz. Goreniya i vzryva* **6** 125 (1986); Pyaternev S V, Pershin S V, Dremin A N *Combustion Explosion and Shock Waves* **22** 756 (1986)

79. Kurdyumov A V, Britun V F, Borimchuk N I, Yarosh V.V. *Martensitnie i diffusionnie prevrasheniya v uglerode in nitride bora pri udarnom szhatii* (*Maternsitic and diffusive transformations in carbon and boron nitride at shock compression*) (2005)

80. Shenderova O et al., *J.Phys.Chem. C*, **115**, 14014 (2011)

81. Dolmatov V Yu. *Uspekhi khimii* **76**(4) 375 (2007); Dolmatov V Yu, *Russ. Chem. Rev.*, **76**(4), 339 (2007)

82. Marks N A, Lattemann M, McKenzie D R, *Phys. Rev. Lett.* **108** 075503 (2012)

83. Deryagin B V, Fedoseev D V *Rost almaza I grafita is gazovoi fazy* (*Diamond and graphite growth from gas phase*) (Moscow, Nauka) 1977.

84. Ashfold M N R et al. *Chem Soc Rev.* 21 (1994)

85. Schwander M, Partes K *Diam.Relat.Mater* **20** 1287 (2011)

86. Rebrov A K *Usp. Fiz. Nauk* **187** 193 (2017); Rebrov A K *Phys. Usp.* **60**, 179 (2017).

87. Khmel'nitskii R A, *Usp. Fiz. Nauk* **185** 143 (2015); Khmel'nitskii R A *Phys. Usp.* **58**, 134 (2015).

88. Palnichenko A V et al. *Nature* **402** 162 (1999)

89. Fukunaga K et al. *Nature* **328** 141 (1987)

90. Nagashima K, Nara M, Matsuda J *Meteorit. Planet. Sci.* **47** 1728 (2012)

91. Bachmann P K, Leers D, Lydtin H *Diam. Relat. Mater.* **1** 1 (1991)

92. Eaton S C, Sunkara M K *Diam. Relat. Mater.* **9** 1320 (2000)

93. Yoshimoto M et al. *Nature* **399** 340 (1999)

94. Hirai T et al. *J Cryst. Growth* **310** 1015 (2008)

95. Popov V T, Polak L S, Fedoseev D V *Kolloidnii Zh.* **49(3)** 618 (1987); Popov V T, Polak L S, Fedoseev D V *Colloid J USSR* **49** 546 (1987)

96. Buerki P R, Leutwyler *J.Appl.Phys.* **69** 3739 (1991)

97. Dubinchuk V T et al., *Doklady AN SSSR* **231**(4) 973 (1976); Dubinchuk V T et al., *Doklady Akad. Nauk SSSR Earth Sci. Sect.* **231** 114 (1976)

98. Daulton T L, Ozima M *Science* **271** 1260 (1996)

99. Daulton T L et al. *Nucl. Instrum. Meth. B* **175-177** 12 (2001)

100. Banhart F, Ajayan P M *Nature* **382** 433 (1996)

101. Wesolowski P et al., *Appl. Phys. Lett.* **71**, 1948 (1997)

102. Byakov V M, Pimenov G G, Stepanova O P *Pis'ma Astron. Zh.* **16**(11) 1051 (1990); Byakov V M, Pimenov G G, Stepanova O P *Sov. Astron. Lett.* **16**, 452 (1990).

103. Nuth J A, Allen J E *Astrophys. Space Sci.* **196**, 117 (1992)

104. Kouchi A et al. *Astrophys. J* **626** L129 (2005)

105. Butenko Yu V et al. *Diam. Relat. Mater.* **17** 962 (2008)

106. Shiryaev A A et al. *Sci. Reps.* **8** 5099 (2018)

107. Fogg J L et al. *Nucl. Inst. Meth. B* **453** 32 (2019)

108. Zaitsev A M Optical Properties of Diamond (Berlin-Hedelberg: Springer, 2001)

109. Dischler B Handbook of Spectral Lines in Diamond (Berlin-Hedelberg: Springer, 2012)

110. Lax M, Burstein E *Phys. Rev.* **97** 39 (1955)

111. Klyuev Yu A et al., *Doklady AN SSSR* **240** 1104 (1978); Klyuev Yu A et al., *Sov. Phys. Dokl.* **23**, 370 (1978)

112. Shiryaev A A et al. *Meteorit. Planet. Sci.* 57(3) 730 (2022)

113. Clark C D, Collins A T, Woods G S in: *The properties of natural and synthetic diamond* (Ed J E Field) (London: Academic Press, 1992) p. 35

114. Stachel T et al., *Rev. Miner. Geochem.,* **88** 809 (2022)

115. De Corte K et al., *Geochim.Cosmochim.Acta* **62** 3765 (1998)

116. Cartigny P et al. *Science* **304** 853 (2004)

117. Vlasov I I et al. in: *Ultrananocrystalline Diamond* (Eds O Shenderova, D M Gruen) (Elsevier, 2012), p. 291

118. Clark C D, Davey S T *J Phys. C: Solid State Phys.* **17** 1127 (1984)

119. Bokii G B, Bezrukov G N, Klyuev Yu A, Naletov A M, Nepsha V I *Prirodnie I sinteticheskie almazy* (*Natural and synthetic diamonds*) (Moscow, Nauka, 1986)

120. Orlinskii S B et al., *Nanosci.Nanotech.Lett.* **3** 1 (2011)

121. Briddon P R, Jones R. *Phisica B* **185** 179 (1993)

122. Titkov S V et al., *Geol. Geofiz.* №1-2 455 (2015); Titkov S V et al., *Russian Geology and Geophysics* **56** 354 (2015)

123. Turner S in: *Nanodiamonds* (Ed. J-C Arnault) (Elsevier, 2017), p. 57

124. Birman J L in: *Theory of Crystal Space Groups and Lattice Dynamics* (Springer, Berlin, Heidelberg, 1974) p. 1

125. Nadolinny V, Komarovskih A, Palyanov Y *Crystals* **7**, 237 (2017)

126. Reutskii V N et al., *Geokhimiya* **55**(11) 1003 (2017); Reutskii V N et al., *Geochemistry International*, **55**(11), 988 (2017)

127. Goss J P et al. *J Phys.: Condens. Mat.* **26**(14) 145801 (2014)

128. Day M C et al. *Diam.Relat.Mater* **143** 110866 (2024)

129. Jiang T., Xu K. *Carbon* **33** 1663 (1995)

130. Kulakova I I *Fizika Tverdogo Tela* **46**(4), 621 (2004); Kulakova I I *Phys. Solid State* 46, 636–643 (2004)

131. Buchatskaya Y et al. *Radiochim. Acta* **103(3)** 205 (2015)

132. Kudryavtsev O S et al. *Nano Letters* **22** 2589 (2022)

133. Ekimov E A et al *Nanomaterials* **12**(3) 351 (2022)

134. Hill H G M et al. *Meteorit. Planet. Sci.* **32** 713 (1997)

135. Andersen A C et al. *Astron. Astrophys.* **330** 1080 (1998)

136. Mutschke H et al. *Astrophys. J* **454** L157 (1995)

137. Mutschke H et al. *Astron. Astrophys.* **423** 983 (2004)

138. Jones A P et al. *Astron. Astrophys.* **416** 235 (2004)

139. Khomich A A et al. *Laser Phys. Lett.* **14**(2) 025702 (2017)

140. Clark C D et al. *Phys. Rev.* **B51**(23) 16681 (1995)

141. Ekimov E A et al., *Diam.Relat.Mater.* **93** 75 (2019)

142. Ekimov E A, Kondrin M V, *Usp. Fiz.Nauk* **187** 577 (2017); Ekimov E A, Kondrin M V *Phys. Usp.* **60** 539 (2017)

143. Zaitsev A M *Mat. Sci. Eng.* **B11** 179 (1992)

144. Zaitsev A M *Phys. Rev.* **B61**(19), 12909 (2000)

145. Gippius A A, et al. *Physica* **116B**, 187 (1983)

146. Tkachev V D, Zaitsev A M, Tkachev V V *Phys. Stat. sol.* **B129**, 129 (1985)

147. Bradac C et al. *Nature Nanotech.* **5** 345 (2010)

148. Vlasov I I et al. *Small* **6** 687 (2010)

149. Terada D et al., *ACS Nano* **13**(6), 6461 (2019)

150. So F T-K et al., *J.Phys.Chem. C* **126(11)** 5206 (2022)

151. Smith B R, Gruber D, Plakhotnik T *Diamond Relat. Mater.* **19** 314 (2010)

152. Bogdanov et al., *Materials*, **15**, 3589 (2022)

153. Vlasov et al. *Adv.Mater.* **21** 808 (2009)

154. Vlasov I I et al. *Nature Nanotech.* **9** 54 (2014)

155. Makino Y et al. *Diamond Relat. Mater.* **112** 108248 (2022)

156. Makino Y et al. *Phys.Stat.Sol. A* 2200342 (2022)

157. Chang H-C et al. *J. Phys. Chem.* **99** 11081 (1995)

158. Cheng C-L et al. *Phys. Rev. Lett.*, **78**, 3713 (1997)

159. Van Kerckhoven C, Tielens A G G M, Waelkens C *Astron. Astrophys.* **384** 568 (2002)

160. Blades J C, Whittet D C B *Monthly Notices Roy. Astron.Soc.* **191** 701 (1980)

161. Aitken D K, Roche PF *Monthly Notices Roy. Astron. Soc.* **196**, 39 (1981)

162. Baas F. et al., *Astrophys. J.* **265** 290 (1983)

163. Geballe T R et al., *Astrophys. J.* **340** L29 (1989)

164. van der Zwet G P et al., *Astron. Astrophys.* **145** 262 (1985)

165. Roche P F, Allen D A, Bailey J A, *Monthly Notices Roy. Astron. Soc* **220**,7 (1986)

166. Acke B, van den Ancker M E *Astron. Astrophys.* **457** 171 (2006)

167. Habart E et al. *Astrophys. J* **614** L129 (2004)

168. Topalovic R et al. *Mon. Not. Roy. Astron. Soc.* **372** 1299 (2006)

169. Goto M et al. *Astrophys. J* **693** 610 (2009)

170. Shiryaev A A et al. *Meteorit. Planet. Sci.* **50**(6) 1005 (2015)

171. Kagi H, Takahashi K, Masuda A *Naturwissenschaften* **77** 531 (1990)

172. Abdullahi I M et al. *Carbon* **164** 442 (2020)

173. Fisenko et al., *Geokhimiya* **43(5)**, 677 (1995); Fisenko et al., *Geochemistry International* **33**(1), 145 (1996)

174. Fisenko et al., *Astron. Vestnik* **31**(1) 82 (1997); Fisenko et al., *Solar System Research*, **31**, 73 (1997)

175. Borchi E et al. *J Phys D – Appl. Phys.* **31** 609 (1998)

176. Cho J-G, Yi B-Y, Kim T-K *Korean J Med. Phys.* **12**(1) 1 (2001)

177. Simonia I A, Mikailov Kh M *Astronomical J* **50**(12) 960 (2006); Simonia I A, Mikailov Kh M *Astron. Rep.* **50** 960 (2006)

178. Chang H-C, Chen K, Chang S K *Astrophys. J* **639** L63 (2006)

179. Chang H-C *J Phys.: Conf. Ser.* **728** 062004 (2016)

180. Darbon S, Perrin J-M, Sivan J-P *Astron. Astrophys.* **348** 990 (1999)

181. Ekimov E A, Lebed Yu B, Kondrin M V *Carbon* **171** 634 (2021)

182. Iakoubovskii K, Adriaenssens G J, Nesladek M *J Phys. Cond. Mat.* **12** 189 (2000)

183. D'Haenens-Johansson et al. *Phys.Rev.B* **84** 245208 (2011)

184. Deshko Y, Gorokhovsky A. *Low Temp. Phys.* **36**, 465 (2010)

185. Shiryaev A A et al. *Phys. Chem. Chem. Phys.* **23** 21729 (2021)

186. Cherchneff I et al. *Astron. Astrophys.* **357**, 572 (2000)

187. Cherchneff I *EAS Publications Series*, **60** 175-184 (2013)

188. Ott U et al., *Publ. Astron. Soc. Australia* **29**, 90 (2012)

189. Dworetsky M M, Persaud J L, Patel K *Mon. Not. Roy. Astron. Soc.* **385**(3) 1523 (2008)

190. Castelli F, Hubrig S, *Astron. Astrophys.* **475** 1041 (2007)

191. Alvarez E et al. *Physica Scripta* **20** 141 (1979)

192. Shiryaev A A et al., in: *Modern stellar Astronomy-2018* (Eds O Yu Malkov, A S Rastorguev, N N Samus', V N Obridko) (IZMIRAN) p. 311, DOI: 10.31361/eaas.2018-1.072

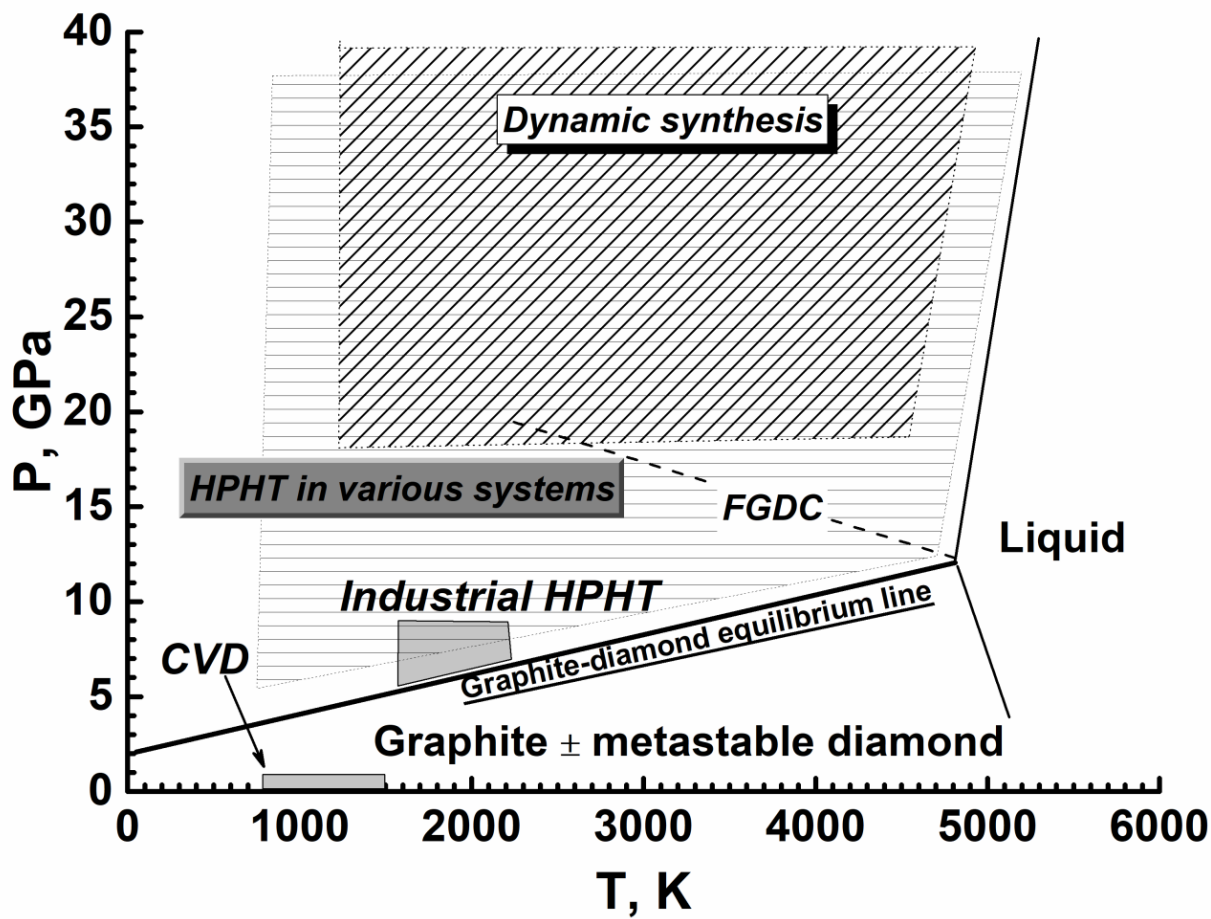

Fig. 1. Schematic carbon phase diagram and regions of diamond synthesis using different methods (adapted from [73, 74]). FGDC – line of solid-state graphite → diamond transformation faster than 1 ms (Fast graphite – diamond conversion); HPHT – industrial diamond synthesis at static *PT* parameters.

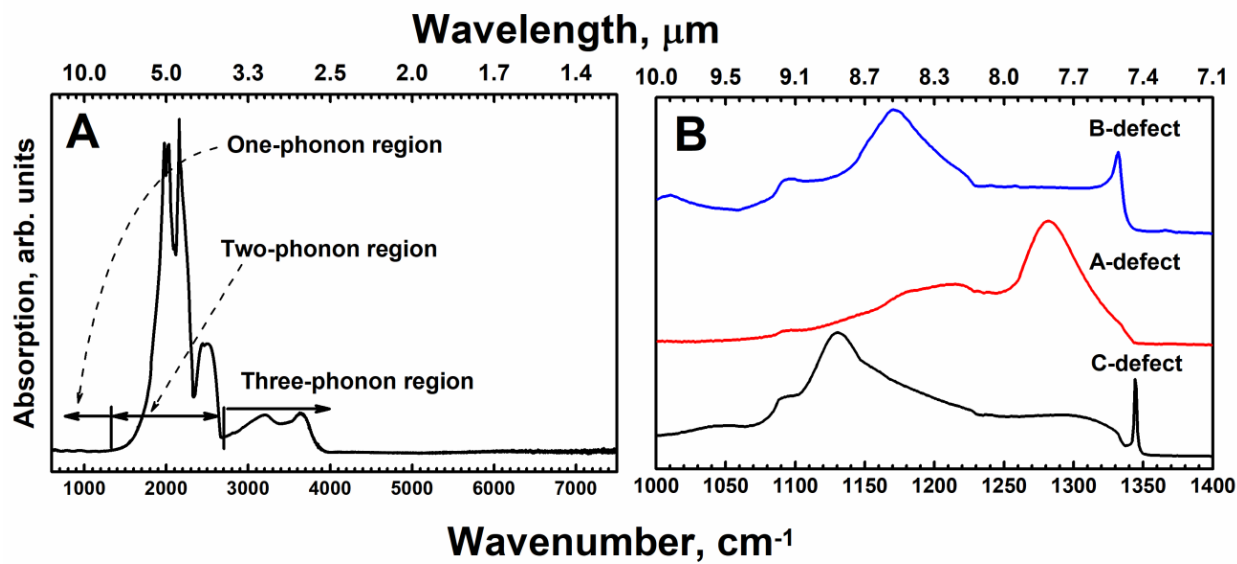

Fig. 2. Absorption spectra of macrodiamond in infrared range. A – nitrogen-less diamond (type IIa). B – spectra of principal nitrogen defects in macrocrystals.

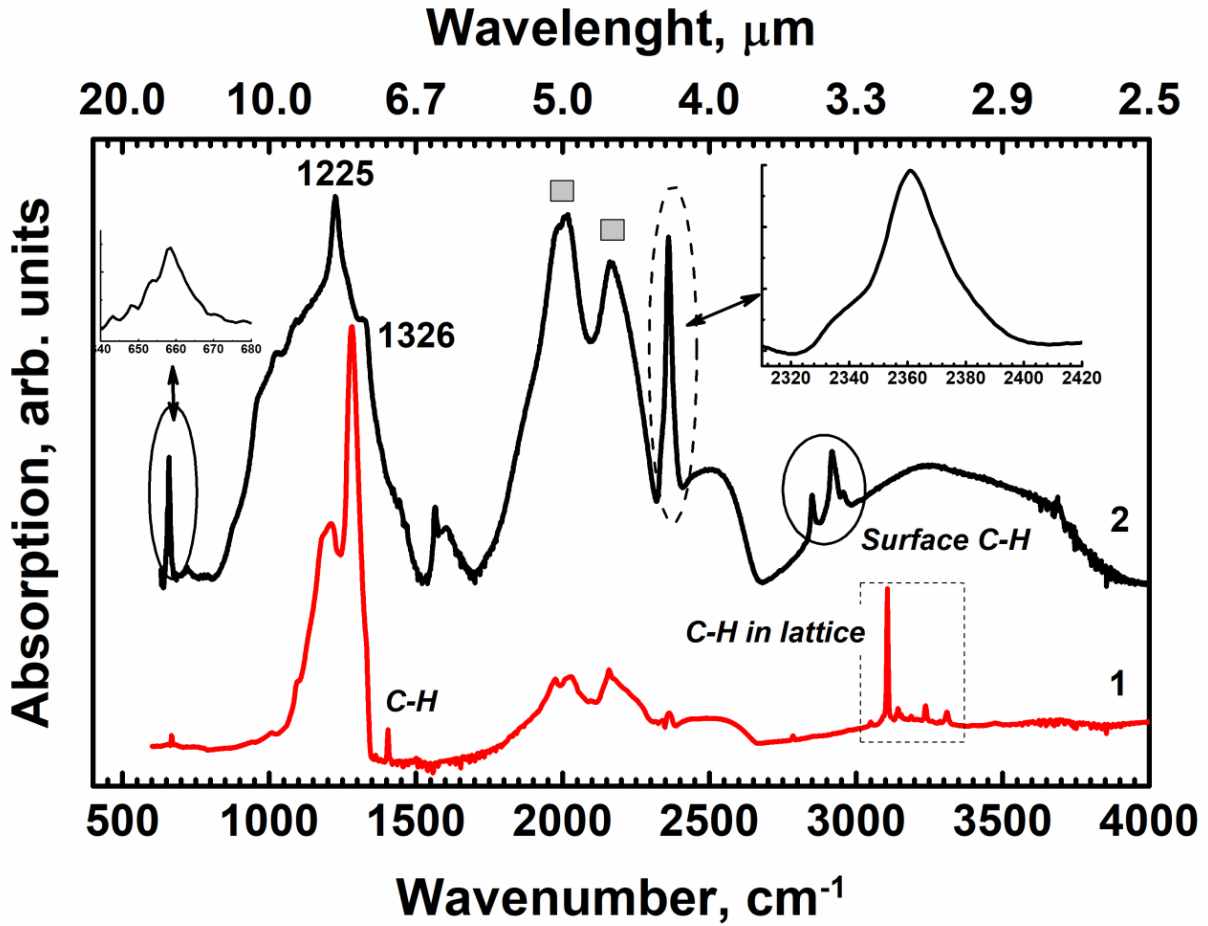

Fig. 3. Infrared spectra of natural macrodiamonds. Curve 1 –specimen with high content of nitrogen in A-form. Vibrations of lattice-bound hydrogen are marked (lines 1405, 3107 cm$^{-1}$ and some other). Curve 2 – bulk nanocrystalline impact diamond (yakutite) [112]. Filled squares mark distorted two-phonon bands. In the one-phonon region, a "triangular" band, presumably related to extended defects, is observed, see text for detail. Peaks of C-H vibrations due to surface contamination are shown. Insets show absorption bands of solid $CO_2$ (phase $CO_2$-I) residing in inclusions.

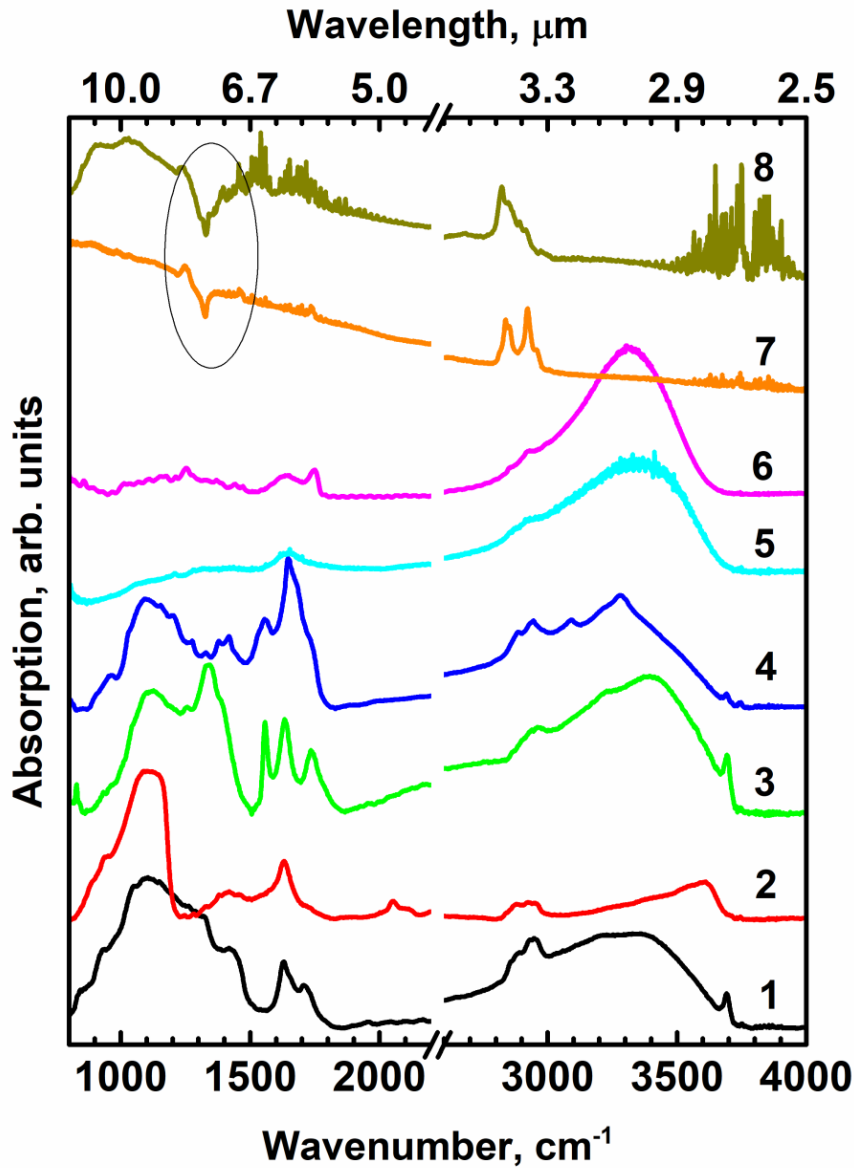

Fig. 4. Infrared absorption spectra of nanodiamonds (features due to atmospheric $CO_2$ are removed). Absorption by diamond lattice in the two-phonon region is not observed. Curves 1-4 – disperse nanodiamonds with various surface functional groups modeified by thermochemical treatment (oxidation, reduction, etc., [131]). Curves 5, 6 – bulk nanocrystalline diamond obtained by powder sintering at high PT-conditions [47]. Spectra in one-phonon region differ across the sample due to PT-gradient during synthesis and gas-phase transport in the pressure cell. Curves 7, 8 – synthetic nanodiamonds with hydrogenated surface showing "transmission window" near the diamond Raman frequency (marked by an ellipse) due to Fano-resonance [132, 133]. Sloped background is also due to the resonance.

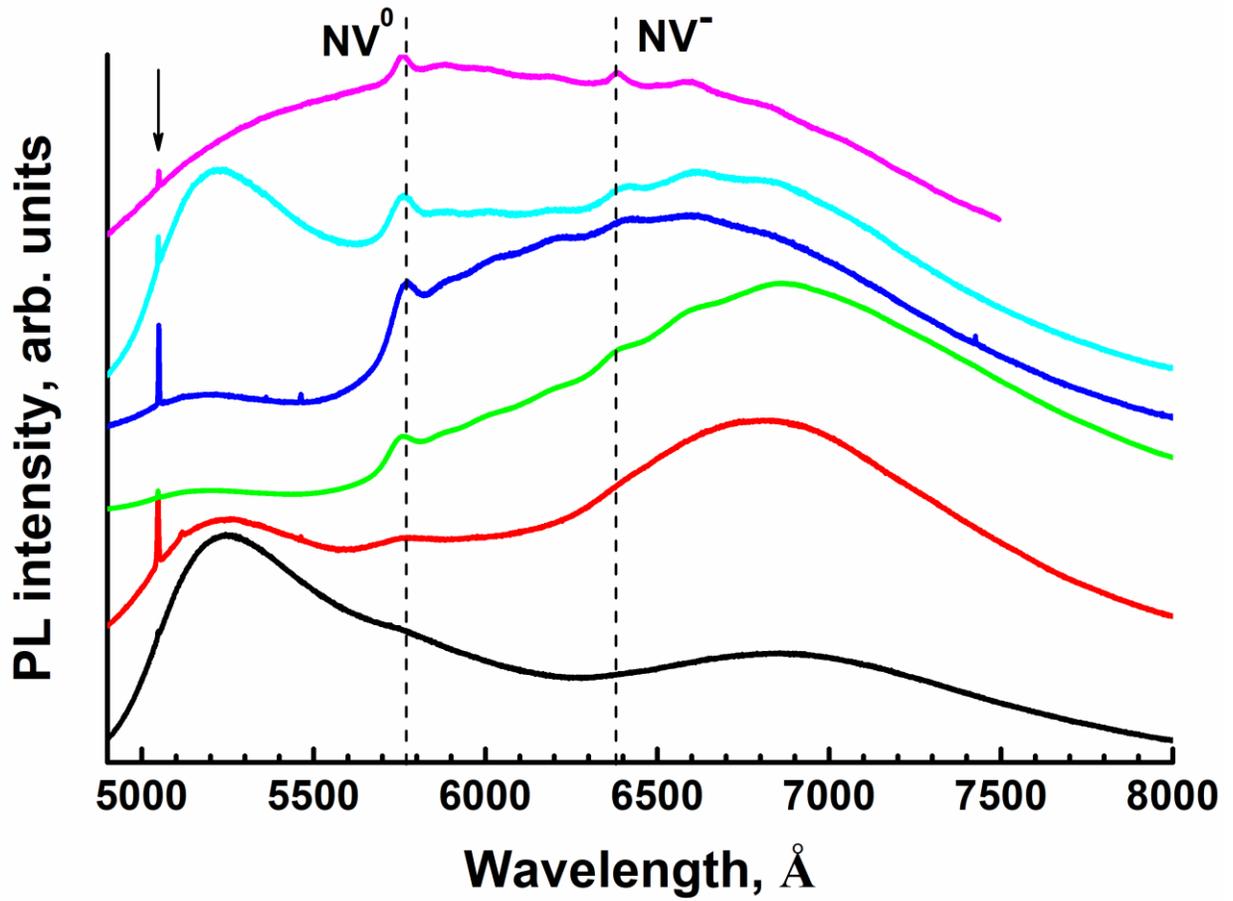

Fig. 5. Room-temperature photoluminescence spectra of diamonds from ureilites ($\lambda_{exc}$=488 nm) [22]. Curves correspond to diamonds from different meteorites. Features of NV defects in various charge states are indicated. Arrow points to diamond Raman peak.

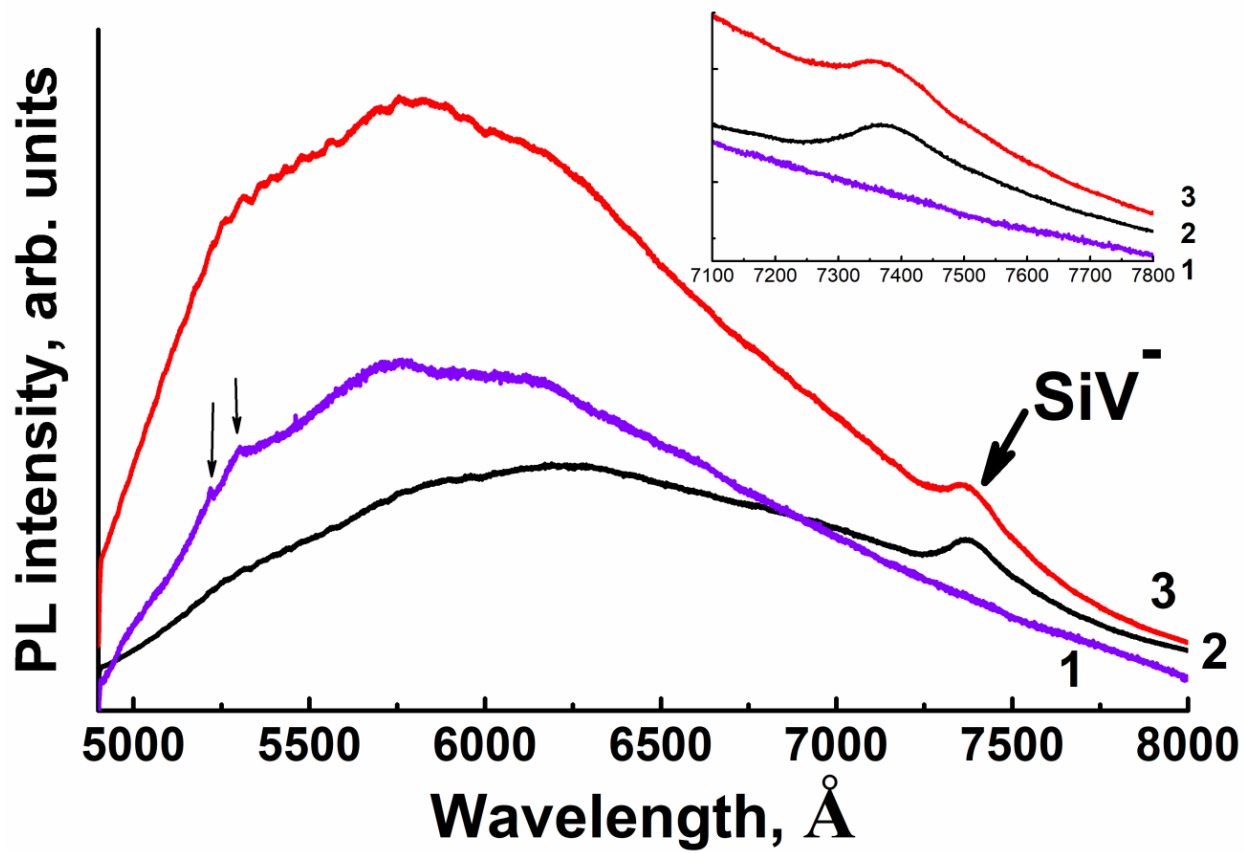

Fig. 6. Room-temperature photoluminescence spectra of disperse nanodiamonds ($\lambda_{exc}$=488 nm) [40]. Curve 1 – deeply purified synthetic nanodiamond; arrows show Raman peaks of surficial sp²-C. Curves 2, 3 – meteoritic nanodiamonds. Band of the SiV⁻ defect is shown; the inset shows this band in detail.